\documentclass[12pt]{article}
\usepackage{amsmath}
\usepackage{euscript}
\usepackage[T1]{fontenc}
\usepackage[latin1]{inputenc}
\usepackage{graphics}
\usepackage{longtable}
\usepackage{color}
\usepackage{pslatex}
\usepackage{subfigure}
\usepackage{pstricks,epsfig,amssymb,axodraw}
\usepackage{cite}  
\begin{document}

\def\lb{\nextline}
 
\def\Order#1{{\cal O}($#1$)}
\def\Kuhn{K\"uhn}
\def\alphapi{\Bigl({\alpha\over\pi}\Bigr)}
\def\sovem{{s\over m^2_e}}
\def\Born{{\rm Born}}
\def\nubar{\bar{\nu}}
\def\nubarnu{\bar{\nu}\nu}
\def\nubart{\bar{\nu}_\tau}
\def\sstrut{$\strut\atop\strut$}

  \def\PL #1 #2 #3 {Phys. Lett. {\bf#1}           (#3)  #2}
  \def\NP #1 #2 #3 {Nucl. Phys. {\bf#1}           (#3)  #2}
  \def\PR #1 #2 #3 {Phys. Rev. {\bf#1}            (#3)  #2}
  \def\PP #1 #2 #3 {Phys. Rep. {\bf#1}            (#3)  #2}
  \def\PRL #1 #2 #3 {Phys. Rev. Lett. {\bf#1}     (#3)  #2}
  \def\CPC #1 #2 #3 {Comp. Phys. Commun. {\bf#1}  (#3)  #2}
  \def\ANN #1 #2 #3 {Annals of Phys. {\bf#1}      (#3)  #2}
  \def\APP #1 #2 #3 {Acta Phys. Pol. {\bf#1}      (#3)  #2}
  \def\ZP  #1 #2 #3 {Z. Phys. {\bf#1}             (#3)  #2}

\def\uncatcodespecials{\def\do##1{\catcode`##1=12 }\dospecials}
\def\setupverbatim{\tt
  \def\par{\leavevmode\endgraf} \catcode`\`=\active
  \obeylines \uncatcodespecials \obeyspaces \parindent=5mm \parskip=0pt}
{\obeyspaces\global\let =\ } 
{\catcode`\`=\active \gdef`{\relax\lq}}
\def\beginverbatim{\par\begingroup\setupverbatim\doverbatim}
{\catcode`\|=0 \catcode`\\=12 
  |obeylines|gdef|doverbatim^^M#1\endverbatim{#1|endgroup}}
\def\Was{\hbox{W\c as\;}}
\def\M{\hbox{\cal M}}
\def\lips{\hbox{Lips}}
\def\Im{\hbox{Im}}
\def\GeV{\hbox{GeV}}
\def\Maj{M_{R}}  \def\Gaj{\Gamma_{R}}
\def\beq{\begin{equation}} 
\def\eqiv{\sim}
\def\eeq{\end{equation}} 
\def\eps{\epsilon}
\begin{titlepage}
 
\begin{flushright} CERN-PH-TH/2006-111, \\
{ IFJPAN-IV-2006-5} 
\end{flushright}
 
\vspace{2cm}
\begin{center}
{\bf\Large
Scalar QED, NLO and PHOTOS Monte Carlo. 
}\end{center}
 
\begin{center}
   {\bf  G. Nanava$^{\dag}$ }\\
   {\em  
       Institute of Nuclear Physics, PAN,
        Krak\'ow, ul. Radzikowskiego 152, Poland
        \\ (On leave from IHEP, TSU, Tbilisi, Georgia)
}\\
     and \\ {\bf  Z. W\c{a}s }\\
   {\em  
       CERN, 1211 Geneva 23, Switzerland\\
        and \\
        Institute of Nuclear Physics, PAN,
        Krak\'ow, ul. Radzikowskiego 152, Poland}\\

\end{center}
\vspace{.2 cm}
\begin{center}
{\bf   ABSTRACT  }
\end{center} 

Recently, QED bremsstrahlung in  $B$-meson decays
 into pair of scalars ($\pi$'s and/or $K$'s) is of  interest.
If experimental acceptance must be taken into account,
 PHOTOS Monte Carlo is often used in experimental simulations. 
 We will use scalar QED to benchmark PHOTOS, even though this theory
 is of limited use for complex objects.  
 We present the analytical form of the kernel used in the older versions 
of PHOTOS, and the new, exact (scalar QED) one.
  Matrix element and phase-space Jacobians are separated in the final weight 
and
future extensions based on measurable electromagnetic form-factors are 
thus possible. 

The  massive phase-space is controlled in the  the program 
 with no approximations. 
 Thanks to the iterative  solution  all leading and 
next to leading 
logarithmic terms are properly reproduced by the Monte Carlo simulation. 
 Simultaneously,  
full differential distributions over complete multiple body phase-space  are provided.
 An agreement of better than 0.01\%  with independent calculations of scalar QED is demonstrated.

 \vspace{.3cm}
\begin{flushleft}
{  CERN-PH-TH/2006-111, \\  IFJPAN-IV-2006-5 \\
 July, 2006}
\end{flushleft}
 
\vspace*{1mm}
\bigskip
\footnoterule
\noindent
{\footnotesize \noindent
Supported in part  by the EU grant MTKD-CT-2004-510126, 
in partnership with the CERN Physics Department,
and the Polish State Committee for Scientific Research 
(KBN) grant 2 P03B 091 27 for years 2004-2006.
$^{\dag}${This work is
partly supported by the EU grant mTkd-CT-2004-510126 in partnership with the
CERN Physics Department and by the Polish Ministry of Scientific Research and
Information Technology grant No 620/E-77/6.PRUE/DIE 188/2005-2008.}
}
\end{titlepage}
\vspace{.5cm}
\begin{center}
{\bf 1. INTRODUCTION}
\end{center}
\vspace{.5cm}

In the analysis of data from high-energy physics experiments, one tries to 
resolve the ``{\it experiment = theory}'' equation.  This non-trivial task
requires that a lot of different effects be considered simultaneously.
From the experimental side, these are mainly detector acceptance and cuts,
which are dictated by the construction and physical properties of the detector. 
The shapes of distributions may be distorted by, say, misidentification 
and residual background contamination. These effects need to be discriminated 
in an appropriate and well-controlled way.
From the theoretical side, {\it all} effects of known physics have to be included in 
predictions as well. Only then experimental data and theoretical predictions can be 
confronted to determine numerical values of  coupling constants or 
effects of new physics (to be discovered). 

A well-defined class of theoretical effects consists of QED radiative 
corrections.
PHOTOS is a universal Monte Carlo algorithm  that simulates the effects of QED 
radiative corrections in decays of particles and resonances.
It is a project with a rather long history: the first version was
released in 1991 \cite{Barberio:1990ms}, followed by version 2.0 
\cite{Barberio:1994qi} (double emission and threshold terms for fermions). 
The package is in wide use \cite{Dobbs:2004qw};
it was applied as a precision simulation tool for the
$W$ mass measurement at the Tevatron \cite{Abazov:2003sv}
and LEP  \cite{Abbiendi:2003jh,Abdallah:2003xn},
and for CKM  matrix measurements 
in decays of $K$ and $B$ resonances 
(NA48 \cite{Lai:2004bt}, KTeV\cite{Alexopoulos:2004up} , Belle \cite{Limosani:2005pi},
BaBar \cite{Aubert:2004te} and  Fermilab \cite{Link:2004vk}).  
Discussion of the different components of systematic errors in PHOTOS is thus of interest.

Throughout the years the core algorithm for the generation of $O(\alpha)$ corrections
did not change much, however, its precision, applicability to various 
processes, and numerical stability improved significantly. 
New functionalities, such as multiple photon radiation
and interference effects for all possible decays were  introduced 
\cite{Golonka:2005dn,Golonka:2005pn}. Recently, the complete first order matrix element was
introduced into PHOTOS for $Z$ decays and complete NLO\footnote{In the
paper, we will use abbreviations  NLO, NLL, NNLO, NNLL to
denote next to leading order, next to leading logarithm,
 next-next to leading order, next-next to leading logarithm
 corrections
with respect to leading order (that is without QED at all).
Meaning for
such abbreviations can not be defined restrictively on the basis
of  approximations used in 
phase-space.
 The properties of the matrix elements need to be specified 
as well. Nonetheless,
 we will use the abbreviations later in the paper to denote such
approximations in the phase-space Jacobians which do not prevent 
the appropriate precision to hold, if  proper choice of matrix elements
is  made as well.} 
 multiple photon predictions for 
that channel were demonstrated to work well \cite{Golonka:2006tw}.
 
Increasing interest in the algorithm expressed by experimental
collaborations (including future LHC experiments and precise measurements for 
$B$ decays) was a motivation to
perform a more detailed study of the potential and precision of the PHOTOS
algorithm. This paper is devoted to the decay of $B$-mesons into a pair of 
scalars. It is a continuation of the previous paper \cite{Golonka:2006tw}  devoted
to $Z$ decays. 
We concentrate our attention on exact phase-space parametrization as used in PHOTOS,
and on the explicit separation of the final weight into parts responsible for the 
following: 
(i) {\it mass dependent} phase-space Jacobians,
(ii) matrix elements and (iii) pre-sampler for peaks.

Simplifications introduced in the matrix element normally used 
 in these scalar B meson decays are removed, the exact kernel of first order
 scalar QED calculation is installed. Such improvement opens the way
 to include data dependent form-factors into matrix elements of PHOTOS, and
physically better results than of 
scalar QED alone.

Our study of the PHOTOS algorithm can be understood as 
another step  in the  on-going effort to find 
 practical solutions of the improved expansions. 
The solution can be seen as a rearrangement 
of the QED perturbation expansion, but this time for the interaction of charged 
scalars with photons and in the case where ultrarelativistic approximations are not valid.

To test PHOTOS we have used predictions of the SANC\cite{Bardin:2005xx} Monte Carlo algorithm. 
SANC is able to calculate the exact first order scalar QED matrix elements
for decays of $B$-mesons into scalars, and covers the full phase space of decay products without any approximations.
Events provided by SANC MC are unweighted.
SANC is a network client-server system for the semi-automatic calculation of Electroweak, QCD and QED radiative corrections at a one-loop
precision level for various  processes(-decays) of elementary particle interactions.

The paper is organized as follows. 
Section 2 is devoted to the description of our results obtained from scalar QED, which will be used later 
in tests and in construction of kernel for single photon emission.
In Section 3 the main properties used in the design of PHOTOS  are presented. In particular, the construction of the weight (NLO level) 
necessary to introduce the complete
first order matrix element is explained in all detail.
 The phase space parametrization used in
the iterative solution of PHOTOS is  given.
Details are collected in the Appendix. 
On the other hand,  quite essential  for complete NLL, 
aspects of the construction
 will still remain poorly documented. That is, how the 
parts of  the single photon emission matrix element are used 
in the (iterated and thus extended to the mutiphoton emissions) kernel.
 Section 4 is devoted to results of numerical tests performed at fixed, first order of QED. Some 
numerical results, obtained with   multi-photon version of the program will be 
collected there for reference.
Finally, section 5 summarizes the paper.

\vspace{.5cm}
\begin{center}
{\bf 2. Scalar QED and B decays }
\end{center}
\vspace{.3cm}

In this section we give the formulae needed in the construction of the 
Monte Carlo routine of two-particle $B$-meson decays and the analytical results of
the decay rates at $\cal{O}(\alpha)$ in scalar-QED when the masses of the decay products are neglected.
The one-loop QED corrections to the width of the decays $B^{0,-} \to H^{-}_{_1} H^{+,0}_{_2}$, where $H_{_{1,2}}$ denotes scalar(pseudo-scalar) particles,
can be represented as a sum of the Born contribution and the contributions due to virtual loop diagrams and soft and hard photon emissions. 
Both virtual and soft contributions factorize to the Born one.
\begin{eqnarray}
      d\Gamma^{\rm Total} = d\Gamma^{\rm Born} \left[ 1 + \frac{\alpha}{\pi}\left(\delta^{\rm soft} + \delta^{\rm virt} \right)\right] + d\Gamma^{\rm\, Hard}\,.
     \label{width_total}
\end{eqnarray}
 Here $d\Gamma^{\rm Born}$ is the tree level differential decay width, $\delta^{virt}$ represents the virtual corrections,
$\delta^{soft}$ denotes the soft photon contribution and  $d\Gamma^{\rm\, Hard}$ is the hard photon contribution.
The Born level distribution in the rest frame of the decaying meson can be written as
\begin{eqnarray}
  d\Gamma^{\rm Born} = \frac{1}{2\,M}\,|A^{\rm Born}|^2\,dLips_2(P \to k_1,k_2)\,,
\end{eqnarray}
where $M$ is the mass of decaying particle, $k_{1,2}$ denote the momenta of decay products,
$A^{\rm Born}$ stands for the corresponding tree level amplitude and $dLips_2(P \to k_1,k_2)$ is the two body differential phase space.
 For the latter we choose the following parametrization 
\begin{eqnarray}
         dLips_2(P \to k_1,k_2) = \frac{1}{32\pi^2}\frac{\lambda^{1/2}(M^2,m^2_{_1},m^2_{_2})}{M^2} d\cos\theta_{1}d\phi_{1}\,,
\end{eqnarray}
where angles $\theta_{1}$ and $\phi_{1}$  define the orientation of momentum $k_{_1}$ in the rest frame of the $B$-meson. 
 As usual we define
$\lambda^{1\over 2}(a, b,c)=\sqrt{a^2+b^2+c^2-2ab-2ac-2bc}$. 
In the case of neutral $B$-meson decay channels $B^{0} \to H^{-}_{_1} H^{+}_{_2}$, 
the scalar-QED calculations for the virtual and soft factors in formula (\ref{width_total}) gives
\begin{eqnarray}
 \delta^{\rm virt} &=& \left[1 + \frac{M^2 - m^2_1 - m^2_2}{\Lambda} \ln\frac{2m_{_1}m_{_2}}{M^2-m^2_1-m^2_2+\Lambda} \right]
                  \ln\frac{M^2}{m^2_\gamma} + \frac{3}{2} \ln\frac{\mu^2_{_{UV}}}{M^2}
                                                                \nonumber\\
              &+& \frac{M^2-m^2_1-m^2_2}{2\Lambda}\left[  {\rm Li_2}\left(\frac{M^2+m^2_1-m^2_2+\Lambda}{2\Lambda}\right) 
                                                        -{\rm Li_2}\left(\frac{-M^2+m^2_2-m^2_1+\Lambda}{2\Lambda}\right)
                                                   \right.      \nonumber\\
                                                      && \phantom{xxxxxxxxxxxxxx}\left.+\; 2\ln\frac{2 M m_{1}}{M^2+m^2_ 1-m^2_2+\Lambda}
                                                                                            \ln\frac{m_{1}\Lambda}{M^3}+(1\leftrightarrow2)+\pi^2
                                                \right]
                                                                \nonumber\\
              &-& \frac{\Lambda}{2M^2}\ln\frac{2m_{_1} m_{_2}}{M^2-m^2_1-m^2_2+\Lambda} + \frac{m^2_2-m^2_1}{4 M^2}\ln\frac{m^2_2}{m^2_1}
                - \frac{1}{2}\ln\frac{m_{_1} m_{_2}}{M^2} + 1;
  \label{virt_nc}
\end{eqnarray}
\begin{eqnarray}
 \delta^{\rm soft} &=& \left[1 + \frac{M^2 - m^2_1 - m^2_2}{\Lambda} \ln\frac{2m_{_1}m_{_2}}{M^2-m^2_1-m^2_2+\Lambda} \right]
                  \ln\frac{m^2_\gamma}{4\omega^2}
                                                                \nonumber\\
              &+& \frac{M^2-m^2_1-m^2_2}{2\Lambda}\left[  {\rm Li_2}\left( \frac{-2\Lambda}{M^2+m^2_1-m^2_2-\Lambda}\right)
                                                        -{\rm Li_2}\left( \frac{2\Lambda}{M^2+m^2_1-m^2_2+\Lambda}\right) + (1\leftrightarrow2)
                                                \right]
                                                                \nonumber\\
              &-& \frac{M^2+m^2_1-m^2_2}{\Lambda} \ln\frac{2 M m_{_1}}{M^2+m^2_1-m^2_2+\Lambda} - (1\leftrightarrow2),
  \label{soft_nc}
\end{eqnarray}
where $m_{_{1,2}}$ are the final meson masses, an auxiliary small parameter $\omega\,$($\ll M/2$) separates the soft and hard photon contributions, 
and $\mu_{_{UV}}$ denotes the  ultraviolet scale. An auxiliary photon mass $m_\gamma$ is used as a regulator of the infrared divergences. 
The ultraviolet singularities are regularized by means of the dimensional regularization. 
We renormalize the wave functions of the external scalar fields in the on-shell scheme, and the point-like weak coupling in the $\rm \overline{MS}$ scheme.
$\Lambda = \lambda^{1/2}(M^2,m^2_1,m^2_2)$ and  $\displaystyle {\rm Li}_2(z)= - \int^z_0\frac{dy}{y}\ln|1-y|$ .

The hard photon distribution $d\Gamma^{\rm\, Hard}$ in scalar-QED  can be expressed as follows
\begin{eqnarray}
   d\Gamma^{\rm\, Hard} = \frac{1}{2\,M}|A^{\rm Born}|^2 4\pi\alpha\left(q_1\frac{k_1.\epsilon}{k_1.k_{\gamma}}-q_2\frac{k_2.\epsilon}{k_2.k_{\gamma}} \right)^2
                        dLips_3(P\rightarrow k_1,k_2,k_{\gamma}).
\label{Hard-two}
\end{eqnarray}
 Here, $q_{1,2}$ are the charges of final mesons, and $k_{\gamma}$ and $\epsilon_{\mu}$ are the photon momentum and  polarization vector respectively.
The three body differential phase space of the decay products, $dLips_3(P\rightarrow k_1,k_2,k_{\gamma})$,  is parametrized, in a rather standard way (see e.g.
\cite{FOWL}),
 as follows:
\begin{eqnarray}
         dLips_3(P\rightarrow k_1,k_2,k_{\gamma}) = \frac{\lambda^{1/2}(1-2E_{\gamma}/M,m^2_{_1}/M^2,m^2_{_2}/M^2)}{16(2\pi)^{5}\;(1-2E_{\gamma}/M)}
                                                   E_{\gamma}dE_{\gamma}\,d\cos\theta_{\gamma}d\phi_{\gamma}\,d\cos\theta^{R}_{1}d\phi^{R}_{1}\,, 
\label{three-lisp} 
\end{eqnarray}
where the angles $\theta^{R}_{1}$ and $\phi^{R}_{1}$ define the orientation of momentum $k_1$ in the rest frame of $(k_1+k_2)$; the photon energy $E_{\gamma}$
and the angles $\theta_{\gamma}$ and $\phi_{\gamma}$,
 that define the orientation of the photon momentum, are given in the rest frame of the decaying particle. 
These parameters vary in the limits: $0 \leq \theta_{\gamma},\,\theta^{R}_{1} \leq\pi \; ,\;\; 0\leq \phi_{\gamma},\,\phi^{R}_{1} \leq2\pi$
 and $\omega \leq E_{\gamma} \leq (M^2-(m_1+m_2)^2)/2M$. The formulae (\ref{width_total})\,--\,(\ref{three-lisp}) are used in the construction 
of the Monte Carlo simulator of the decays under consideration. An analytical integration in (\ref{Hard-two}) over the phase space variables (\ref{three-lisp})
can be done easily. Below we give the result of integration in the massless limit of the final mesons (i.e. $m_{_1},m_{_2} \equiv m\rightarrow 0$)
because of its simplicity:
\begin{eqnarray}
    \Gamma^{\rm\, Hard} = \Gamma^{\rm Born} \frac{\alpha}{\pi} \left[  \left(1-\ln\frac{m^2}{M^2}\right)\ln\frac{4\omega^2}{M^2}
                                                                   - 2\ln\frac{m^2}{M^2}
                                                                   - \frac{\pi^2}{3} + 4
                                                           \right].
    \label{hard_lim_nc}
\end{eqnarray}
The virtual correction depends on the ultraviolet scale $\mu_{_{UV}}$, which should cancel in the total decay width
because of the scale dependence of the point-like weak coupling. The infrared divergence cancels out in the sum of virtual 
and soft contributions, as it must. The total decay width, which is the sum of the contributions (\ref{virt_nc}), (\ref{soft_nc}),
and (\ref{hard_lim_nc}), is also free of  $\omega$ and 
of the final meson mass singularity in accordance with the KLN theorem ~\cite{Kinoshita:1962ur}--\cite{Lee:1964is}:
\begin{eqnarray}
 \Gamma^{\rm Total} = \Gamma^{\rm Born} \left[1 + \frac{\alpha}{\pi}\left(\frac{3}{2}\ln\frac{\mu^2_{_{UV}}}{M^2} + 5\right)\right].
  \label{total_lim_nc}
\end{eqnarray}

The same calculations can be done for the charged $B$-meson decay channels  $B^{-} \to H^{-}_{_1} H^{0}_{_2}$. In this case for various
contributions in formula (\ref{width_total}) we obtain:
\\[2mm]
-- Virtual photon contribution
\begin{eqnarray}
 \delta^{\rm virt} &=& \left[1 + \frac{M^2 + m^2_1 - m^2_2}{\Lambda} \ln\frac{2M m_{_1}}{M^2+m^2_1-m^2_2+\Lambda} \right]
                  \ln\frac{M m_{_1}}{m^2_\gamma} + \frac{3}{2} \ln\frac{\mu^2_{_{UV}}}{M m_{_1}}
                                                                \nonumber\\
              &+& \frac{M^2+m^2_1-m^2_2}{2\Lambda}\left[  
                                                         {\rm Li_2}\left( \frac{M^2-m^2_1-m^2_2+\Lambda}{2\Lambda}\right) 
                                                        -{\rm Li_2}\left( \frac{M^2-m^2_1-m^2_2-\Lambda}{-2\Lambda}\right)
                                                  \right.       \nonumber\\ && \phantom{xxxxxxxxxxxx}\left.
                                                      +\;{\rm Li_2}\left( \frac{M^2+m^2_2-m^2_1-\Lambda}{-2\Lambda}\right) 
                                                        -{\rm Li_2}\left( \frac{M^2+m^2_2-m^2_1+\Lambda}{2\Lambda}\right)                                                         
                                                  \right.       \nonumber\\
                                                     && \phantom{xxxxxxxxxxxx}\left.+\; 2\ln \frac{2 M m_{_1}}{M^2+m^2_1-m^2_2+\Lambda}
                                                                                         \ln\frac{\Lambda}{M m_{_2}}
                                                                                        -\ln \frac{2 M m_{_2}}{M^2+m^2_2-m^2_1+\Lambda}
                                                                                         \ln\frac{M^2}{m^2_{1}}
                                                \right]
                                                                \nonumber\\
              &+& \frac{\Lambda}{2m^2_{2}} \ln\frac{2M m_{_1}}{M^2+m^2_1-m^2_2+\Lambda} - \frac{M^2-m^2_1}{4 m^2_{2}}\ln\frac{m^2_1}{M^2} + 1;
  \label{virt_cc}
\end{eqnarray}
-- Soft photon contribution
\begin{eqnarray}
 \delta^{\rm soft} &=& \left[1 + \frac{M^2 + m^2_1 - m^2_2}{\Lambda} \ln\frac{2 M m_{_1}}{M^2+m^2_1-m^2_2+\Lambda} \right]
                  \ln\frac{m^2_\gamma}{4\omega^2}
                                                                \nonumber\\
              &+& \frac{M^2+m^2_1-m^2_2}{2\Lambda}\left[  {\rm Li_2}\left( \frac{-2\Lambda}{M^2+m^2_1-m^2_2-\Lambda}\right)
                                                        -{\rm Li_2}\left( \frac{2\Lambda}{M^2+m^2_1-m^2_2+\Lambda}\right) 
                                                \right]
                                                                \nonumber\\
              &-& \frac{M^2+m^2_1-m^2_2}{2\Lambda} \ln\frac{2 M m_{_1}}{M^2+m^2_1-m^2_2+\Lambda};
  \label{soft_cc}
\end{eqnarray}
-- Hard photon contribution
\vspace{-1mm}
\begin{eqnarray}
   d\Gamma^{\rm\, Hard} &=&  \frac{1}{2\,M} |A^{\rm Born}|^2 4\pi\alpha\left(q_1\frac{k_1.\epsilon}{k_1.k_{\gamma}}-q\frac{P.\epsilon}{P.k_{\gamma}} \right)^2
                        dLips_3(P\rightarrow k_1,k_2,k_{\gamma}), \nonumber \\[4mm]
    \Gamma^{\rm\, Hard} &=& \Gamma^{\rm Born} \frac{\alpha}{\pi} \left[ \left(1+\frac{1}{2}\ln\frac{m^2}{M^2}\right)\ln\frac{4\omega^2}{M^2}
                                                                  + \ln\frac{m^2}{M^2}
                                                                  - \frac{\pi^2}{6} + 3
                                                           \right].
\label{Hard-one}
\end{eqnarray}
Again after  integration  over the phase space variables, the  massless limit of the final mesons 
(i.e. $m_{_1},m_{_2} \equiv m\rightarrow 0$) was used in the last formula.
Finally, summing contributions (\ref{virt_cc}), (\ref{soft_cc}) and (\ref{Hard-one}), we obtain the following expression
for the total decay width:
\begin{eqnarray}
      \Gamma^{\rm Total} &=& \Gamma^{\rm\, Born} \left[1 + \frac{\alpha}{\pi}\left(\frac{3}{2}\ln\frac{\mu^2_{_{UV}}}{M^2} - \frac{\pi^2}{3} + \frac{11}{2}\right)\right].
  \label{total_lim_cc}
\end{eqnarray}

We have checked that the factors $\delta^{\rm soft}$ and $\delta^{\rm virt}$ in formula (\ref{width_total}), for both charged and neutral $B$-meson decays, 
provide the same numerical results as the corresponding expressions in ~\cite{Baracchini:2005wp}.

To be assured of the accuracy of SANC Monte Carlo integration (which is a by-pro\-duct of MC simulation), 
we compared Monte Carlo results with  the analytical calculations of the total decay rate. The result of this comparison is shown in 
Table 1\footnote{Please note that these numbers are for the purpose
of our test only, the overall $B-H-H$ coupling constants do not match the experimental data.}.
The agreement is thus better than $10^{-4}$ in this test, where mass effects were included.
\begin{table}[h]
\begin{center}
\begin{tabular}{|c|c|c|}
\hline
Channel              & $\Gamma^{\rm Total}_{\rm AC}$, $10^{-3}$MeV & $\Gamma^{\rm Total}_{\rm MC}$, $10^{-3}$MeV \\[1mm]
\hline
$B^- \to \pi^-\pi^0$ &  0.373629                                &  0.3736(4)                               \\
$B^- \to K^- K^0$    &  0.367586                                &  0.3675(9)                               \\
\hline 
$B^0 \to \pi^-\pi^+$ &  0.377392                                &  0.3773(8)                               \\
$B^0 \to K^- K^+ $   &  0.371414                                &  0.3714(2)                               \\
\hline
\end{tabular}
\end{center}
\vspace*{-3mm}
\caption{\it Comparison of the $B$-meson total decay widths produced by analytical calculations (second column) and by Monte Carlo (third column).
 The last significant digit of the Monte Carlo results is given in bracket.
}
\end{table}

\newpage

\vspace{.5cm}
\begin{center}
{\bf 3. Exact phase-space and matrix element.}
\end{center}
\vspace{.5cm}

To start any discussion of the implementation of complete first order QED radiative corrections in $B$ decay, one has to specify the parametrization of the 
complete phase-space  slots of the fixed final state multiplicity.

 Let us start with the explicit expression for the parametrization of 
an  $n+1$ body phase-space
in decay of  the object of four-momentum $P$\;  ($P^2=M^2$), as used in PHOTOS Monte Carlo. 
As our aim is to define iterative relations, let us denote the four momenta 
of the first  $n$ decay products as $k_i$ ($i=1,n$) and the last $n+1$ decay product as $k_{n+1}$.
In our case  the $n+1$-th particle will  always be the real and massless  
photon\footnote{However the construction does not rely on a photon to be massless. 
In principle it can be 
applied to define other  phase space relations, for example the
emission of an extra massive pion or  emission of a pair of heavy particles.}. 
In the later steps of our construction the masslessnes of  photons and properties 
of QED matrix elements will be used. 

In the following, notation from refs. \cite{Was:1994kg,Jadach:1993hs} will be used. 
We will not rely on any particular results of these papers.
We only point to other,  similar options for
the exact $n$-body phase-space parametrizations, which are also in use.

The Lorentz invariant phase-space is defined as follows:
\begin{eqnarray}
dLips_{n+1}(P) &=&
{d^3k_1 \over 2k_1^0 (2\pi)^3}\; . . .\;{d^3k_n \over 2k_n^0 (2\pi)^3}
{d^3k_{n+1} \over 2k_{n+1}^0 (2\pi)^3}
(2\pi)^4 \delta^4\Bigl(P - k_{n+1}- \sum_{i=1}^n k_i\Bigr)\nonumber\\
&=&
d^4p\delta^4(P -p-k_{n+1}){d^3k_{n+1} \over 2k_{n+1}^0 (2\pi)^3}
{d^3k_1 \over 2k_1^0 (2\pi)^3} \;. . .\;{d^3k_n \over 2k_n^0 (2\pi)^3}
(2\pi)^4 \delta^4\Bigl(p -\sum_{i=1}^n k_i\Bigr)\nonumber\\
&=&
d^4p\delta^4(P -p-k_{n+1}){d^3k_{n+1} \over 2k_{n+1}^0 (2\pi)^3} dLips_n(p\to k_1 ... k_n),
\label{Lips_n+1}
\end{eqnarray}
 where extra  integration variables, four vector $p$ 
(compensated with $\delta^4\bigl(p -\sum_1^n k_i\bigr) $) is introduced.
 If further,    $M_{1...n}$ (compensated with 
$\delta\bigl(p^2 -M_{1...n}^2\bigr) $) is introduced,
the element of the phase space takes the form:
\begin{eqnarray}
dLips_{n+1}(P) &=&
{dM_{1...n}^2 \over (2\pi)} dLips_2(P \to p\ k_{n+1}) \times dLips_n(p \to k_1 ... k_n)\nonumber\\
&=&
dM_{1...n}^2  \biggl[d\cos\hat{\theta} d\hat{\phi} {1 \over 8(2\pi)^3}
{\lambda^{1\over 2}(M^2, {M_{1...n}^2 },{m_{n+1}^2 })\over M^2}\biggr]
\times dLips_n(p \to k_1 \dots k_n).\nonumber\\
\label{Lips_n+1.3}
\end{eqnarray}
The part of the phase space Jacobian corresponding to integration over the direction 
and energy of the last particle (or equivalently invariant mass $M_{1...n}$ of the remaining  system
of ${1...n}$ particles) 
is explicitly given.   
We will use later in the formulas   $m_i^2=k_i^2$, and analogously  $M_{i \dots n}$,
defining invariant masses of $k_i \dots k_n$ systems. 
The integration over the angles $\hat{\theta}$ and $\hat{\phi}$ is defined in the $P$ rest-frame. 
The integration over the invariant mass, $M_{1\dots n}$, is limited by phase space boundaries.
Anybody familiar with the phase-space parametrization as used in FOWL \cite{FOWL}, TAUOLA \cite{Jadach:1993hs}, or many other programs will find the above explanation quite standard\footnote{%
%
The  parametrizations of such a type, 
use  properties of the Lorentz group in an explicit manner, in particular 
 measure,  representations and 
their products. That is why, they are  useful, for event building Monte Carlo 
programs   in 
phase-space constructions based 
on boosts and rotations.
}. 

The question of choice of axes with respect to which angles are defined, and order 
in kinematical construction, is less trivial. The choice for the particular option stems from
 necessity to presample collinear singularities.  It is rather well known that the choice of 
the reference directions for the parametrization of the unit sphere is free, and can be used 
to advantage.
We will use related, but  somewhat different freedom of choice. Instead of  variables $\hat{\theta}\; \hat{\phi}$ defining orientation of  $k_{n+1}$ in $P$ rest-frame we will use  angles $\theta_1 \; \phi_1$
orienting $k_1$ (also  in $P$  rest-frame). The Jacobian for this reparametrization of unit sphere equals 1 as well\footnote{Let us point to another
difference with respect to angles  $\theta^{R}_{1}$, $\phi^{R}_{1}$  used 
in  formula (\ref{three-lisp}) 
(and  also in refs. \cite{FOWL},\cite{Jadach:1993hs}). For example, if $n=2$ 
then our $dLips_{n=2}$ phase-space is parametrized  by the angles 
$\theta$ and $\phi$ only. The two angles will define orientation 
of $k_\gamma$ with respect
to frame used for quantization of internal state of $k_1$, this is different
choice  than the one used for $\phi^{R}_{1}$, even though $\theta=\theta^{R}_{1}$. The Jacobian for the corresponding change of variables equals 1 also.}.


 Formula (\ref{Lips_n+1.3}) can 
be iterated and  provide a parametrization of the phase space with an arbitrary number of final state 
particles.  In such a case, the question of orientation of the frames used to define the angles
and the order of $M_{i\dots n}$ integrations (consequently, the choice of limits for $M_{i \dots n}$ integration), 
becomes particularly rich.
Our choice is defined in ref. \cite{Barberio:1994qi}.
We will not elaborate on this  point here, nothing new was introduced
for the purpose of our study.

If the invariant mass $M_{1\dots n}$ is replaced with the photon energy defined in the $P$ rest-frame, $k_\gamma$,
then the phase space formula can be written as:
\begin{eqnarray}
 dLips_{n+1}(P) &=&
\biggl[ k_\gamma dk_\gamma  d\cos\hat{\theta} d\hat{\phi} {1 \over 2(2\pi)^3}
\biggr]
\times dLips_n(p \to k_1 ... k_n),
\label{Lips_n+1.5}
\end{eqnarray}
If we would have $l$  photons accompanying $n$ other particles,
then the factor in square brackets is iterated.
The statistical factor ${1 \over l!}$ would complete the form of the phase space 
parametrization, similar to the formal expansion
of the exponent.
The last formula, supplemented with definition of frames with 
respect to which angles are defined  is used
to define the full kinematic configuration of the event. From angles and energies ($k_{\gamma_i})$ 
of photons and also angles, energies and masses for other decay products, 
four-momenta of all final state particles can be constructed.

If in formula (\ref{Lips_n+1.5}) instead of  $dLips_n(p \to k_1 ... k_n)$  one would use 
 $dLips_n(P \to k_1 ... k_n)$  the {\it \bf tangent space} would be obtained.  
Then $k_{n+1}$ photon does not affect  
 other particles' momenta at all, and thus has no boundaries on energy or direction. If this
formula would be iterated then
all such photons would be  independent from one another as well\footnote{ Expression (\ref{Lips_n+1.5}) would be  slightly more complicated if instead of photons
a massive particle was to be added.}.
Energy and momentum constraints on the photon(s) are introduced with  the relation between tangent 
and real $n+1$-body phase-space. The formula defining one step in the iteration reads 
as follows\footnote{The $ \{ \bar k_1,\dots,\bar k_{n}\}$ can be identified with the event before 
the radiation of $k_\gamma$ is introduced.}:
\begin{eqnarray}
 && dLips_{n+1}(P\to k_1 ...  k_n,k_{n+1})=  dLips_{n}^{+1\; tangent} \times W^{n+1}_n, \nonumber\\[3mm]
&&dLips_{n}^{+1\; tangent} = dk_\gamma d\cos\theta d\phi \times dLips_n(P \to \bar k_1 ... \bar k_n),
\nonumber \\
&&\{k_1,\dots,k_{n+1}\} = {\bf T}\bigl(k_\gamma,\theta,\phi,\{\bar k_1,\dots,\bar k_n\}\bigr).
\label{Jacobians}
\end{eqnarray}
  The 
 $W^{n+1}_n$ depends on details of ${\bf T}$, and will be thus 
given later in formula~(\ref{Wnn}). To justify (\ref{Jacobians}), 
we have to convolute formula (\ref{Lips_n+1.3})
for $Lips_{n+1}(P \to  k_1 ... k_n,k_{n+1})$  with itself (for $Lips_{n}(p \to k_1 ... k_n)$):
\begin{eqnarray}
Lips_{n+1}(P \to k_1 ... k_n,k_{n+1}) &=& {dM_{1\dots n} ^2 \over 2\pi}  Lips_{2}(P \to k_{n+1} p) \times Lips_{n}(p \to  k_1 ... k_n) \nonumber \\ 
Lips_{n}(p \to k_1 ... k_n) &=& {dM_{2\dots n}^2 \over 2\pi}  Lips_{2}(p \to k_1 p') \times Lips_{n-1}(p' \to  k_2 ... k_n)
\label{AA}
\end{eqnarray}
and use it also for $Lips_{n}(P \to  \bar k_1 ... \bar k_n)$:
\begin{eqnarray}
Lips_{n}(P \to  \bar k_1 ... \bar k_n) &=& {dM_{2\dots n}^2 \over 2\pi}  Lips_{2}(P \to \bar k_1 \bar p') \times Lips_{n-1}(\bar p' \to  \bar k_2 ... \bar k_n).
\label{BB}
\end{eqnarray}

 Note that our  tangent space of variables $ dk_\gamma  d\cos{\theta} d{\phi}$ is unbounded from above and the limit is introduced
by $W_n^{n+1}$ which is set to zero for the configuations outside the phase-sace. 
In principle, we should distinguish between variables like $M_{2\dots n} $ for invariant mass
of $k_2 \dots k_n$ and  $\bar M_{2\dots n} $ for invariant mass
of $\bar k_2 \dots \bar k_n$, but in our choice for $G_n$, $G_{n+1}$ below,  
$M_{2\dots n}= \bar M_{2\dots n}$ and  $M_{1\dots n} $ is defined anyway for the $n+1$-body phase space only.

 We direct the reader to refs.\cite{Barberio:1990ms,Barberio:1994qi} 
for an alternative presentation.  Let us remark that 
formula (\ref{Jacobians}) is quite general, many options, motivated by the
properties of the matrix elements, can be introduced. Generally the
transformation $T$ may differ from the choice to choice quite a lot. 
The most straightforward choice can be based on any $n$ and $n+1$ body phase-space 
parametrizations using invariant masses and angles (e.g. exactly as in TAUOLA
\cite{Jadach:1993hs} formulas 11 to 13).

If 
\begin{equation}
G_n \; : \; M_{2\dots n} ^2,\theta_{1},\phi_{1}, M_{3\dots n} ^2,\theta_{2},\phi_{2}, \dots, \theta_{n-1},\phi_{n-1} \;   \to \;\bar k_1 \dots \bar k_n 
\label{G-1}
\end{equation}
and 
\begin{equation} 
G_{n+1} \;:  \; k_\gamma,\theta,\phi,M_{2\dots n}^2,\theta_{1},\phi_{1}, M_{3\dots n} ^2,\theta_{2},\phi_{2},\dots, \theta_{n-1},\phi_{n-1} \;   \to \;k_1 \dots k_n,k_{n+1}
\label{G-2}
\end{equation}
then
\begin{equation}
{\bf T}=G_{n+1}( k_\gamma,\theta,\phi,G_n^{-1}(\bar k_1,\dots,\bar k_n)).
\end{equation}
The ratio of the Jacobians (factors $\lambda^{1/2}$  like in  formula (\ref{Lips_n+1.3}),
 etc.) form the factor $W^{n+1}_n$, which in our case is  rather simple, 
\begin{equation} 
W^{n+1}_n=  {k_\gamma}  {1 \over 2(2\pi)^3}
 \times \frac{\lambda^{1/2}(1,m_1^2/M_{1\dots n}^2,M^2_{2 \dots n}/M_{1\dots n}^2)}{\lambda^{1/2}(1,m_1^2/M^2,M^2_{2\dots n}/M^2)},
\label{Wnn}
\end{equation}
because of  choice for $G$ as explained in the Appendix. Note that 
${k_\gamma}=\frac{M^2-M_{1\dots n}^2}{2M}$.
There are additional benefits from such a choice. In all relations
$\bar k_2= Lk_2$, ...,  $\bar k_n= Lk_n$ and $\bar p'= Lp'$
common Lorentz transformation
 $L$ is used.     Transformation  $L$ 
is defined from $k_1,\bar k_1,\bar p',p'$ and $P$;
 internal relations between four vectors  $k_2 ... k_n$,
($\bar k_2 ... \bar k_n$) are not needed.

Formula (\ref{Jacobians}) can be realized algorithmically in the following way:
\begin{enumerate}
\item
 For any point in n-body phase space (earlier generated event), described for example with the explicit
configuration of four vectors $\bar k_1 ... \bar k_n$,
 coordinate variables can be calculated, using formula (\ref{G-1}).
\item Photon  variables can be generated according to Eq. (\ref{Jacobians}). The weight $W^{n+1}_n$ has to be 
also attributed.
\item Variables obtained in this way from the old configuration and the one of a  photon 
can be used to 
construct the new kinematical configuration for the $n+1$-body final state. 
The phase-space weight, which is zero for configurations outside phase space 
boundaries, can be calculated at this point from (\ref{Jacobians},\ref{Wnn}) and finally combined  with
the matrix element.
\end{enumerate}

 Here we have chosen  two sub-groups of 
particles. The first one consisted of particle 
  1 alone, and the second, of particles 2 to n combined together.
 Obviously in the case of 2-body decays as  discussed in this paper, 
there is not much choice when construction of the first photon is performed.

By iteration, we can generalize formula (\ref{Jacobians}) to the case of $l$  photons and
we write:
\begin{eqnarray}
 && dLips_{n+l}(P\to  k_1 ...  k_n, k_{n+1} ...  k_{n+l})= 
\frac{1}{l!} \prod_{i=1}^l  \biggl[ dk_{\gamma_i}  d\cos\theta_{\gamma_i} d\phi_{\gamma_i} 
 W^{n+i}_{n+i-1}\biggr]
\times dLips_n(P \to \bar k_1 ... \bar k_n),  \nonumber\\ &&
\{k_1,\dots,k_{n+l}\} = {\bf T}\bigl(k_{\gamma_l},\theta_{\gamma_l},\phi_{\gamma_l},{\bf T}\bigl( \dots,
{\bf T}\bigl(k_{\gamma_1},\theta_{\gamma_1},\phi_{\gamma_1},\{\bar k_1,\dots,\bar k_n\}\bigr) \dots\bigr).
\label{barred}
\end{eqnarray}
In this formula we can easily localize the   { \bf tangent space} for the multiple 
photon configuration.  In this space, each photon is independent from  
other particles' momenta.  Note that   
it is also possible  to fix upper boundary on $k_{\gamma_i} $ arbitrary high.
Photons are independent one from another as well. 
Correlations appear later,  thanks to iterated  transformation {\bf T}.
  The factors  $ W^{n+i}_{n+i-1}$ are calculated when 
constraints on each consecutive photon are introduced; the previously constructed ones 
are included in  the $n+i-1$ system\footnote{Configurations of $k_{\gamma_i} $ which can 
not be resolved are replaced with the ones of  that photon dropped out.}.

Of course, for the tangent space to be useful, the choice
of the definition of {\bf T} must be restricted at least by the condition $\{ k_1, \cdots k_n \} \to 
\{ \bar k_1, \cdots \bar k_n \}$ if all $k_{\gamma_i} \to 0$.\footnote{In fact further constraints have to be fulfilled to enable presampling for the collinear
singularities.
Note that  variables 
$k_{\gamma_m},\theta_{\gamma_m},\phi_{\gamma_m}$ are used at a time
of the $m-$th step of iteration  only, and are not needed elsewhere
in construction of the physical phase space; the same is true for invariants
and angles  $M_{2\dots n} ^2,\theta_{1},\phi_{1} ,\dots, \theta_{n-1},\phi_{n-1} \;   \to \;\bar k_1 \dots \bar k_n $ of (\ref{G-1},\ref{G-2}), which are also
redefined at each step of the iteration.
}

It is important to realize that  one has to  choose matrix elements on the tangent 
space to complete the construction used in PHOTOS. The number and energies 
of photons will be generated on the tangent space first.
 Regularization of (at least) soft singularity must be defined.
Rejection, and event construction, is performed with the help of formula
(\ref{Jacobians})  for each consecutive photon. It  diminishes photon 
multiplicity with respect to the one defined for the tangent space.
 Of course, as rejection implements changes in phase-space density, 
a matrix element (with  virtual corrections) of the physical space 
can be introduced  as well.

The treatment of the phase space presented here lies at the heart of the construction 
of PHOTOS kinematics, and was used since its beginning. It exhausts the case when there is
only one charged particle in final state.  For multiple 
charged particle final states new complication appear, because  all
collinear configurations need simultaneous attention, and not only 
the one along $k_1$ direction.
A presampler with multichannel generation is needed.
In our case we follow the same method\footnote{We will omit
details here, because for the two-body final states the  
complications manifest themselves only in the case of multiple photon generation, 
see Appendix. } as explained  in ref. \cite{Jadach:1993hs}.

In the standard version of PHOTOS, as published in \cite{Barberio:1990ms,Barberio:1994qi},
the following matrix element is used for single photon emission when there is
{\it  only one charged particle} in final state:
\begin{eqnarray}
|{\cal M}|^2_{PHOTOS} &=&  |A^{\rm Born}|^2 WT_3^{old} 
\label{X-fotos}
\end{eqnarray}
where
\begin{eqnarray}
 WT_3^{old} &=&  \frac{ 4\pi\alpha }{WT_1WT_2}
\frac{2(1-x)}{1+(1-x)^2}
          \left(1-\frac{m_R^2}{1-\beta^2\cos^2\theta}\right)
 \frac{1+\beta\cos\theta }{2}
 \frac{1 - \sqrt{1-m_R^2} \cos\theta}{1-\beta\cos\theta } \nonumber \\[2mm]
\beta&=& \sqrt{1-4\frac{m_1^2}{M^2(1-x)} \frac{1}{(1-x+ (m_1^2-m_2^2)/M^2)^2}}\nonumber \\
 x&=&\frac{2E_\gamma}{M}\frac{M^2}{M^2-(m_1+m_2)^2},\;\;\;
m_R^2=4 \frac{m_1^2}{M^2(1+m_1^2/M^2)^2}.
\label{WT3-old}
\end{eqnarray}
The old and lengthy  approximation $WT_3^{old}$ for $ WT_3$ implemented in standard PHOTOS is 
 kept for compatibility with ref.~\cite{Barberio:1994qi}.
The expression defining $WT_3$  without approximations reads:
\begin{eqnarray}
|{\cal M}|^2_{exact}&=& |A^{\rm Born}|^2 4\pi\alpha\left(q_1\frac{k_1.\epsilon}{k_1.k_{\gamma}}-q\frac{P.\epsilon}{P.k_ {\gamma}} \right)^2\nonumber \\
      &=& |A^{\rm Born}|^2  4\pi\alpha \; \frac{8}{M^2}  \times
        \frac{\lambda\left(1,\frac{m_1^2}{\tau},\frac{m_2^2}{\tau}\right) \left(1-\frac{2E_\gamma}{M}\right) \sin^2\theta}
         {2\Bigl(\frac{2E_{\gamma}}{M}\Bigr)^{2}\Bigl(1 +\frac{m_1^2-m_2^2}{\tau}-\lambda^{1/2}\left(1,\frac{m_1^2}{\tau},\frac{m_2^2}{\tau}\right)
  \cos\theta\Bigr)^2} \nonumber \\
      &=& |A^{\rm Born}|^2  4\pi\alpha \; \frac{8}{M^2}  \times
 WT_3(P,k_1,k_2,k_\gamma);
\label{X-mustraal}
\end{eqnarray}
here, $(k_1+k_2)^2=\tau$. In both, the standard and exact version of PHOTOS, the same phase-space 
parametrization and presampler for collinear and soft singularities are used.
Together with $ WT_3$ the following factors
related to phase space  $ WT_1$, $ WT_2$, contribute to the  final weight implemented in routine {\tt PHOCOR} of PHOTOS:
\begin{eqnarray}
 WT_1(P,k_1,k_2,k_\gamma) &=& \frac{\lambda^{1/2}\left(1,\frac{m_1^2}{M^2},\frac{m_2^2}{M^2}\right)}
      {\lambda^{1/2}\left(1,\frac{m_1^2}{\tau},\frac{m_2^2}{\tau}\right)}
\frac{2E_\gamma}{M} \nonumber \\
WT_2(P,k_1,k_2,k_\gamma)& =& \frac{2(1- \cos\theta \sqrt{1-m_R^2})}{1+(1-x)^2}
\frac{M}{2E_\gamma}
\label{wtes}
\end{eqnarray}
The expression for $ WT_1$ can be deciphered from formula (\ref{Jacobians}) and  $ WT_2$
is related to presamplers for collinear and soft singularities. The factors  $WT_1$ and $ WT_2$
are used in present paper  in definition of 
$ WT_3^{old}$ (formula \ref{WT3-old}) only.


The combined effect of the virtual and real corrections on the total rate
manifests through
$\frac{\Gamma^{\rm Total}}{\Gamma^{\rm Born}}$. The virtual corrections are
included into PHOTOS through this factor. Let us point that 
the ratio of (\ref{X-mustraal}) and (\ref{X-fotos}) constitutes the basic element of upgrading 
PHOTOS functionality to the complete first order\footnote{ When the option of multiple radiation is used in PHOTOS, 
the single photon emission kernel is iterated. This leads to some complications.}.
The correcting weight can be chosen simply as:
\begin{equation}
wt=\frac{|{\cal M}|^2_{exact}}{|{\cal M}|^2_{PHOTOS}}\frac{\Gamma^{\rm Born}}{\Gamma^{\rm Total}}.
\label{wgt}
\end{equation}
For the standard version of PHOTOS  the virtual corrections 
are required to be such that the total decay rate remains unchanged after
complete QED corrections are included. 
 
In case of final states with two charged particles in PHOTOS the formula (\ref{wgt}) 
need to be modiffied with one of the following versions of the interference weight:
\begin{eqnarray}
wt&=&\sum_{i=1,2}\; \frac{|{\cal M}|^2_{exact}}{|{\cal M}|^2_{PHOTOS}}\Bigg|_i 
 \; \; \frac{\Gamma^{\rm Born}}{\Gamma^{\rm Total}} \; WT_{INT}^i
\nonumber \\
WT_{INT}^i&=&\frac{\left(q_1\frac{k_1.\epsilon}{k_1.k_{\gamma}}-q_2\frac{k_2.\epsilon}{k_2.k_{\gamma}} \right)^2}
{ \left(q_1\frac{k_1.\epsilon}{k_1.k_{\gamma}}-q_1\frac{P.\epsilon}{P.k_{\gamma}} \right)^2
+ \left(q_2\frac{k_2.\epsilon}{k_2.k_{\gamma}}-q_2\frac{P.\epsilon}{P.k_{\gamma}} \right)^2
} \nonumber \\
WT_{INT-option}^i&=& J_{i}\frac{\left(q_1\frac{k_1.\epsilon}{k_1.k_{\gamma}}-q_2\frac{k_2.\epsilon}{k_2.k_{\gamma}} \right)^2}
{ \left(q_1\frac{k_1.\epsilon}{k_1.k_{\gamma}}-q_1\frac{P.\epsilon}{P.k_{\gamma}} \right)^2 J_1
+ \left(q_2\frac{k_2.\epsilon}{k_2.k_{\gamma}}-q_2\frac{P.\epsilon}{P.k_{\gamma}} \right)^2 J_2
}
\nonumber \\
J_1&=& \frac{1}{WT_1(P,k_1,k_2,k_\gamma)WT_2(P,k_1,k_2,k_\gamma)}
\nonumber \\
J_2&=&  \frac{1}{WT_1(P,k_2,k_1,k_\gamma)WT_2(P,k_2,k_1,k_\gamma)}
\label{wgt1}
\end{eqnarray}
The sum over two generation channels $i=1,2$ related to emisson from $q_1$ and $q_2$ is to be performed%
\footnote{Formulae  (\ref{X-mustraal}) and (\ref{X-fotos}) for  $ |{\cal M}|^2_{exact}$ 
and $|{\cal M}|^2_{PHOTOS}$ are for emissions in case of single charge 
final state only, the interference weigt is to introduce the exact matrix 
element for process of two charged scalar final state.}.
The form of $WT_{INT}^i$ results from the exact expressions, formulae (\ref{Hard-one}) and (\ref{Hard-two}).
However, phase space and multichannel presampler specific terms (\ref{wtes}) need to be  discussed. 
Presence of  $J_1$ and $J_2$ in interference weight is optional, but only for single photon radiation. 
The factor $J_{1,2}$ ($J_{1}$ or $J_{2}$) should cancel the $WT_1 \cdot WT_2$ term of the generation branch 
used for this particular event generation. In general, the
 absence of $J_{1,2}$ terms is due to properties of the second order matrix element%
\footnote{For example the form $WT_{INT-option}$ is inappropriate for configurations when the first generated 
photon is hard and the second soft.}. 
  For the time being analogies to the case of $Z$ decay have to be used 
instead of the proof.

Once we have completed the description of our internal correcting weight necessary for PHOTOS 
to work in the NLO regime, we will turn to the numerical results. 
\vspace{.5cm}
\begin{center}
{\bf 4. Results of the  tests }
\end{center}
\vspace{.5cm}

The most attractive property of  Monte Carlo is the possibility to implement selection criteria for the
theoretical predictions that coincide with the experimental ones. Especially in the case of the final 
state bremsstrahlung presence of experimental cut-offs is essential, as they usually significantly 
increase the size of the QED effects. 

In this section we will concentrate however, on the following pseudo-observables, as used in
ref. \cite{Andonov:2002mx,Nanava:2003cg}:
\begin{itemize}
\item
 {\it Photon energy in the decaying particle rest frame:} this
observable is sensitive mainly to the leading-log (i.e. collinear) non-infrared
(i.e. not soft) component of the distributions.
\item
  {\it Energy of the final-state charged particle:} as  the 
previous one, this
observable is sensitive mainly to the leading-log (i.e. collinear) non-infrared
(i.e. not soft) component of the distributions.
\item
  {\it Angle of the photon with  
final-state charged particle:} this
observable is sensitive mainly to the non-collinear  (i.e. non-leading-log) 
but soft 
(i.e. infrared) component of the distributions.
\item
  {\it Acollinearity angle of the final-state scalars:} this
observable is sensitive mainly to the non-collinear  (i.e. non-leading-log) 
and  non-soft 
(i.e. non-infrared) component of the distributions.
\end{itemize}

We will  start our  comparison for   $B^- \to \pi^0 K^-(\gamma)$ 
and  PHOTOS running without improvements from the complete matrix element:
the agreement looks good, see fig. \ref{P0Km_distr_NotCorrected_p}, and  holds 
over the entire range of the distributions, even though densities vary  by up to 8 orders of magnitude. Differences can hardly 
be seen. To visualize the differences,
in fig. \ref{P0Km_ratio_NotCorrected_p}, the ratios of the distributions are plotted. Similar to what was seen in the tests for $Z$ decays 
\cite{Golonka:2006tw} local discrepancies may reach up to 15 \%  for $\cos\theta_{acoll.} > 0.5$. Note however that  those 
regions of the phase-space contribute at the level of $10^{-6}$ to the total decay rate.  
Once the matrix element is switched on, see fig.  \ref{P0Km_ratio_Corrected1_p}, where ratios of distribution
are plotted, the agreement
become excellent, even at  a statistical level of $10^9$ events. It was of no use to repeat
the plots of the distributions with the corrected weight in PHOTOS, as the 
plots could not be distinguished from the ones of fig. \ref{P0Km_distr_NotCorrected_p}.

Encouraged by the excellent performance in the case of the decay into final states with a single charged 
particle, let us now turn to decays into two charged mesons. To avoid accidental 
simplifications, we have selected final states with scalars of different masses
($B^0 \to \pi^- K^+(\gamma)$).

Again, as can be seen from figs. \ref{PmKp_distr_NotCorrected_p} and  \ref{PmKp_ratio_NotCorrected_p},
agreement between PHOTOS using the standard kernel and SANC is rather good, but some differences persist.
Once the complete kernel is switched on, fig.  \ref{PmKp_ratio_corrected_p}, the agreement 
is quite amazing. 
In this case, the interference weight, and the multiple singularity structure of the pre-sampler Jacobians, formula 
(\ref{wgt1}), were tested as well. Both versions $WT_{INT}$ and  $WT_{INT-option}$ 
gave the same results for the case of single photon emission.
 However only the first version,  $WT_{INT}$ turned out to be consistent with exponentiation. 
 To complete the tests 
 for multichannel emissions, final states with more than two massive
decay products need to be studied, preferably for multi-photon radiation as well. 

Let us comment that not only the shapes of the distribution agree in an excellent manner
between PHOTOS and SANC simulations, also the number of events with photons of energy below  the certain threshold
agreed  better than 0.01 \%, thus were consistent with each other within a statistical error of $10^9$ 
event samples. 
The  excellent  agreement, presented in our paper, combined with other results published before,
 help to confirm   that theoretical
effects normally missing in  PHOTOS are small, but if necessary can be introduced into the code. 
It is also important to note that the agreement  provides powerful technical test of the generator.

Finally, let us point out that early  
versions of the program, before 2004, were not  reaching that level of technical sophistication. To establish it
required a major effort.
Kinematical variables used in PHOTOS differ from those of SANC.
The differences could arise due to technical problems, also if for example the Born-level events
which are to be modified by PHOTOS would not fulfill energy-momentum conservation or 
particles momenta were not  on mass-shell,
at the numerical double precision level. This point must always be checked for every new installation 
of PHOTOS in an experimental environment. For that purpose we have collected numerical results, given in Table 2., for
the cumulant of bremsstrahlung decay width: $G(E_{test})=\Gamma(E_{test})/\Gamma^{\rm Total}$, where  $\Gamma(E_{test})$
denotes the decay width, integrated over energy carried by all bremsstrahlung  photons combined
 up to maximum of $E_{test}$. 

\vspace{.5cm}
\begin{center}
{\bf 5. Summary}
\end{center}
\vspace{.5cm} 

This paper was devoted to the study of bremsstrahlung corrections in the decay of $B$-mesons into
pair of scalars of rather large masses. The results  were presented in the analytical
form and  Monte Carlo simulations which were later compared. 

To quantify the size of the Next to Leading order effects normally 
missing in PHOTOS we have installed  into the program
the complete scalar-QED first order expression for the $B$ decay matrix element. 
After modification, the differences
between PHOTOS and the matrix element calculation embodied in SANC
were  below statistical error of $10^9$ events for all of our benchmark distributions.
 Both PHOTOS and SANC were run at fixed first order without exponentiation.
The agreement provides a technical test of the simulations from both of the two 
programs as well.

The improvement of the agreement due to the introduction of a correcting weight 
could come with a price. That was the case with the decay of $Z$. However
because our $B$-mesons are scalar, 
the  complications did  not materialize and a correcting weight 
can be installed to standard PHOTOS versions.
 On the other hand, introduced improvements are numerically 
small. Deficiencies of standard PHOTOS are 
localized  in  corners 
of bremsstrahlung phase space populated by photons of very high energies and angularly  well separated
from final state mesons. Those regions of the phase space  
weigh less than 0.005  to the total
rate and differences in that region approach 20 \% of their size, at most. 
The effects are thus significantly  lower 
than 0.1 \%, if quantified in units of the integrated  $B$ decay rate of a particular channel. 
Also, in those regions, the predictive power of scalar QED
is rather doubtful. That is why we do not think it is
urgent for users
to change the PHOTOS correcting weight  to enable the complete NLO, unless 
 measured 
form-factors become available. Contribution to the systematic error of PHOTOS due to
incompleteness of the old kernel (with respect to scalar QED) does not depend on experimental cuts and is thus of no phenomenological importance for today.

Our paper  was  not only focused on numerical  results due to final state 
bremsstrahlung in $B$ decays. Aspects of mathematical organization 
of the program for calculation of radiative corrections for $B$ production
and decay was discussed as well. Approximations used in PHOTOS affects matrix elements and {\it not} phase
space, which is treated exactly including all mass effects. 

Details of phase-space parametrization and other aspects necessary for implementation of NLO effects are collected
for the first time.
Generation of the phase space starts from  the tangent space constructed from
an eikonal approximation but used also for the hard photons, even of energies 
above the available maximum. In the second step, phase-space constraints are 
enforced. The method  is similar
to  exclusive exponentiation \cite{Jadach:2000ir}.

Complete re-analysis of the final weight for decays into scalars
was presented. Parts corresponding to matrix elements, phase space Jacobians and generator pre-samples
were explicitly separated. Special care was devoted to mass terms.
Analytic form of the single 
photon emission kernel (i. e. matrix element with approximation) used in standard version of PHOTOS,
was also explicitly given. That is why, the analysis presented here can be easily  
 extended to other  
decay channels. It is the first time that we have presented such a study for particles other than 
elementary fermions and in the case where mass terms  of order 
$\frac{\alpha}{\pi} \frac{m_{1,2}^2}{M^2}$ are 
not neglected. Our analytical calculations  agree with the  results of reference 
\cite{Baracchini:2005wp} exactly and could serve as a  basis of our technical tests of the program.

The numerical results collected  in Table 2. can be used as a
 technical test of PHOTOS installation in end-user simulation set-ups. We strongly recommend such tests
to be performed. In these tests the agreement between PHOTOS and 
SANC (or simple semi-analytical expressions for higher order simulations)  was significantly better 
than 0.1 \% for all entries.

In case of program operating for multiple photon radiation, energy momentum constraints
are introduced for each consecutive photon, step by step,  and conformal symmetry is not exploited
in that procedure. Deatails of phase-space parametrization used for multiple photon radiations
were presented. In principle,
for $B$ decays   
and  multiple photon radiation in PHOTOS, a similar level of agreement as in ref. 
\cite{Golonka:2006tw} for $Z$ decay is expected, but the appropriate reference 
distributions do not exist yet. In particular the second order, scalar-QED,
 matrix element for   $B$ decays was not available for us.
That is why we think, that the matrix-element related details of the program construction, 
 necessary for implementation of NLL effects in general case, must remain delegated to the forthcoming work; 
 probably to the times, when we will have at our disposal other second order matrix elements 
than just of $Z \to l \bar l$. 
At present, only dispersed (and for NLL based on analogies
with $Z$ decays only)
results of refs.~\cite{Barberio:1994qi,Was:2004ig,Golonka:2006tw} are available for that purpose.

PHOTOS used for decays of $B$-mesons into scalars provides
an example of program working for multiple emissions from both outgoing charged lines.  
It covers  complete phase space, no special treatment is needed for 
 the hard photon emission regions. Also mass terms have been 
included without any approximations.

On the technical level it is worth mentioning that the NLO correcting 
weight of PHOTOS is used as an internal weight. 
All generated events  remain  weight 1, exactly as it was in the case of $Z \to \mu^+\mu^-$ decay.

In principle, if necessary, even complete higher order matrix elements (NNLO level) could 
 be incorporated with the help of correcting  weights.
This interesting point  definitely goes beyond the scope of the present paper and also beyond the phenomenological
interest for any foreseeable future. This is equally true for the possible
extensions to simulations in QCD, which are also outside the scope of the paper.  

\vskip 3mm

{\bf Acknowledgements.}

Useful discussions with E. Barberio, P. Golonka, Z. Nagy, F. Tkachov and T. Sj\"ostrand are acknowledged. 

\clearpage

\begin{figure}[b]
  \begin{center}
    \includegraphics[ width=160mm,height=270mm, keepaspectratio]{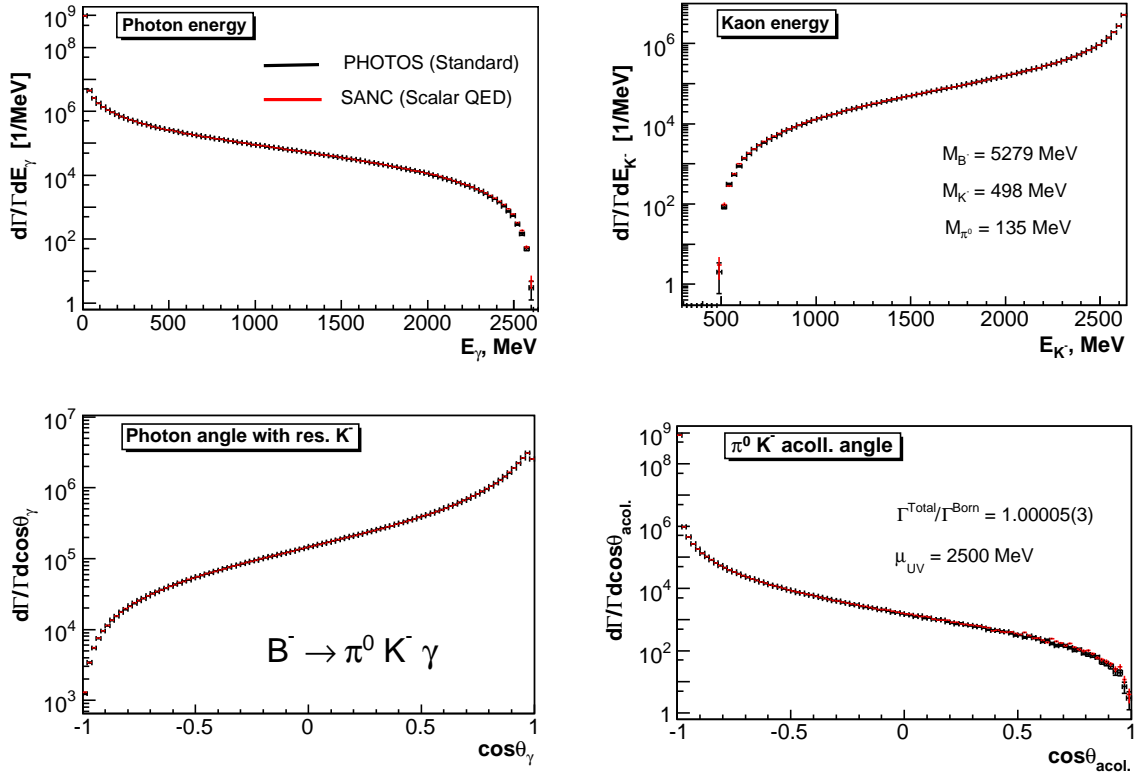}
  \end{center}
  \caption{\label{P0Km_distr_NotCorrected_p} Results from PHOTOS,
    standard version, and SANC for $B^- \to \pi^0 K^-(\gamma)$ decay are 
    superimposed on the consecutive plots. Standard
    distributions, as defined in the text, and logarithmic scales are used.
    The distributions from the two programs overlap almost completely. 
Samples of $10^9$ events were used. 
The ultraviolet scale, $\mu_{_{UV}}$, was chosen to leave total
decay width unchanged by QED.
  }
\end{figure}

\clearpage

\begin{figure}[b]
  \begin{center}
    \includegraphics[ width=160mm,height=270mm, keepaspectratio]{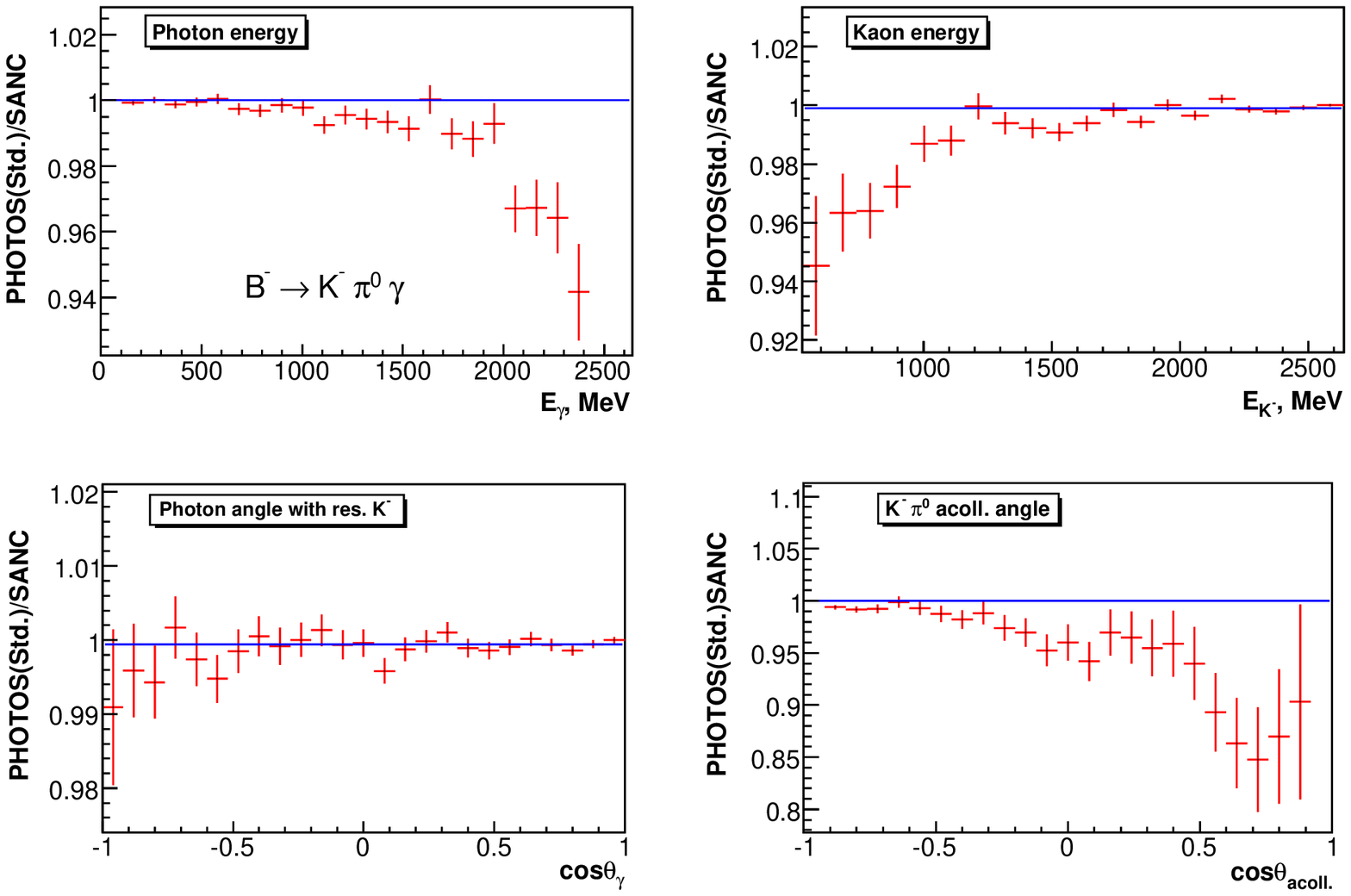}
  \end{center}
  \caption{\label{P0Km_ratio_NotCorrected_p} Results from PHOTOS,
    standard version, and SANC for ratios of the $B^- \to \pi^0 K^-(\gamma)$  distribution in fig.\ref{P0Km_distr_NotCorrected_p} 
    are presented. 
    Differences between PHOTOS and SANC are small, but are clearly visible now.
  }
\end{figure}

\clearpage

\begin{figure}[b]
  \begin{center}
    \includegraphics[ width=160mm,height=270mm, keepaspectratio]{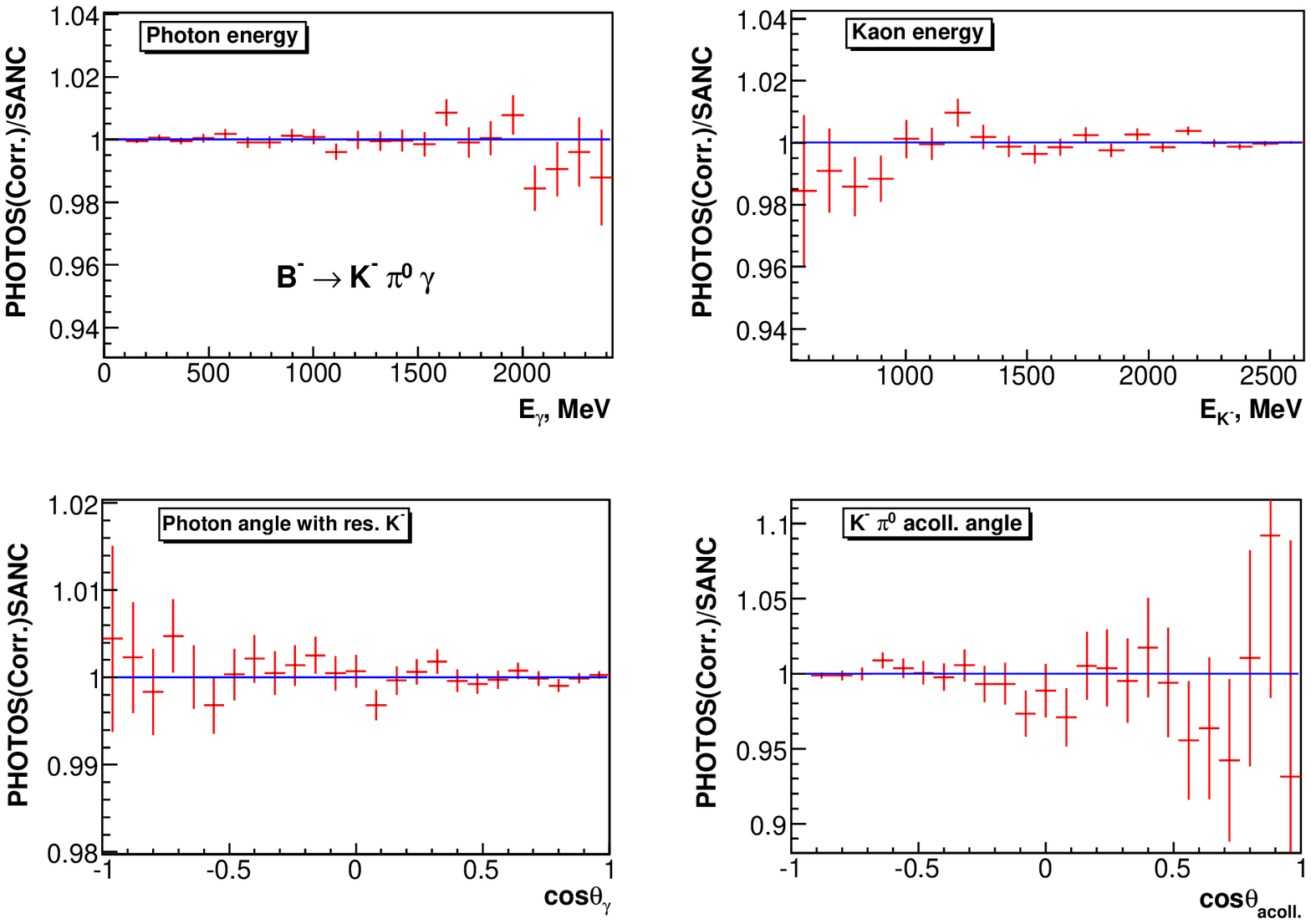}
  \end{center}
  \caption{\label{P0Km_ratio_Corrected1_p} Results from PHOTOS with the exact matrix element, and SANC
    for ratios of the $B^- \to \pi^0 K^-(\gamma)$  distributions. Differences between PHOTOS and SANC are below statistical error for samples of $10^9$ events.
  }
\end{figure}

\clearpage

\begin{figure}[b]
 \begin{center}
   \setlength{\unitlength}{0.1 mm}
   \begin{picture}(105,95)
   \end{picture}
   \includegraphics[ width=160mm,height=270mm, keepaspectratio]{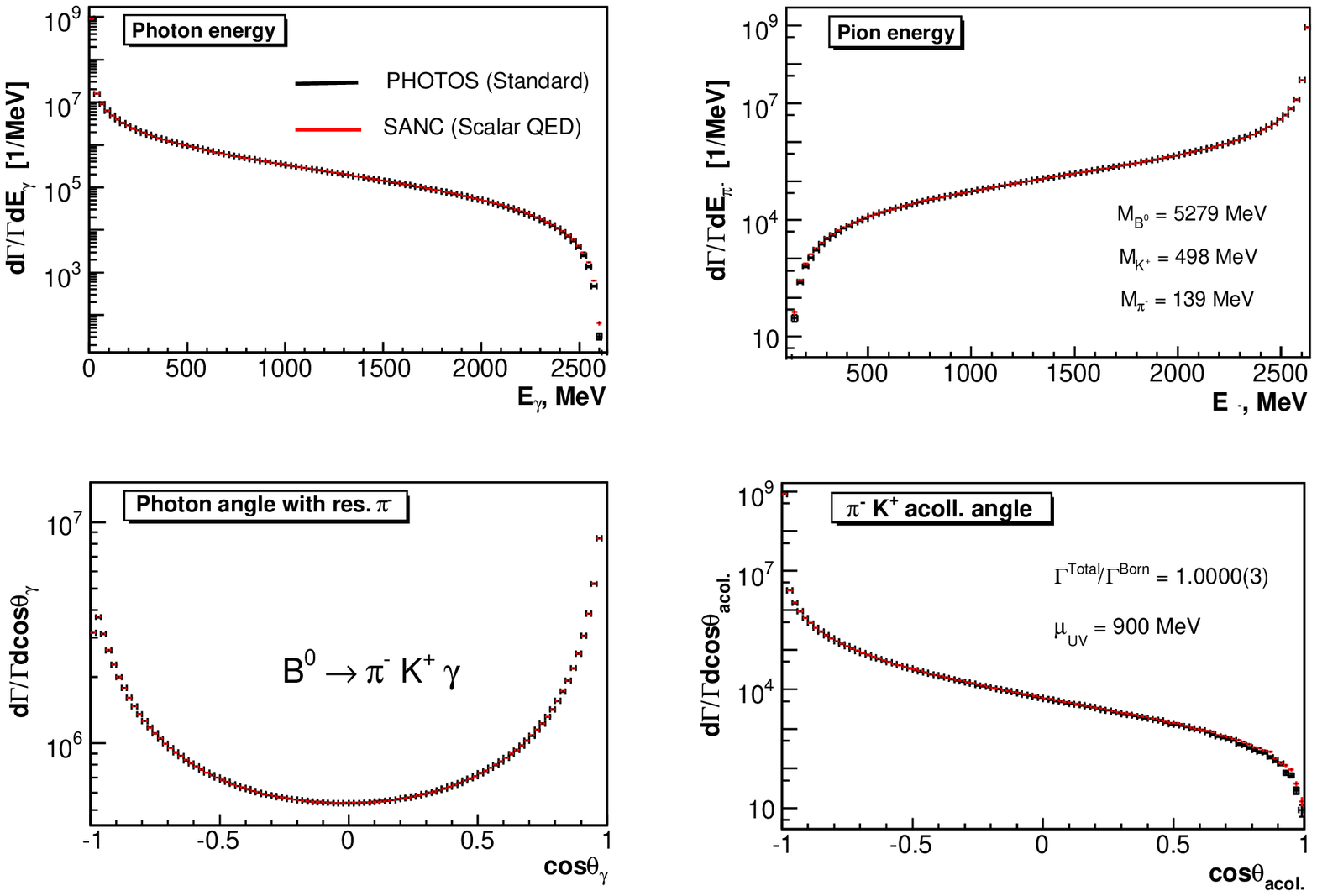}
 \end{center}
   \caption{\label{PmKp_distr_NotCorrected_p} Results from PHOTOS,
standard version, and SANC for $B^0 \to \pi^- K^+(\gamma)$ decay are 
superimposed on the consecutive plots. Standard
distributions, as defined in the text and logarithmic scales are used.
The distributions from the two programs overlap almost completely. 
Samples of $10^9$ events were used.
The  ultraviolet scale, $\mu_{_{UV}}$, was chosen to leave total
decay width unchanged by QED.
}
\end{figure}

\clearpage

\begin{figure}[b]
 \begin{center}
  \includegraphics[ width=160mm,height=270mm, keepaspectratio]{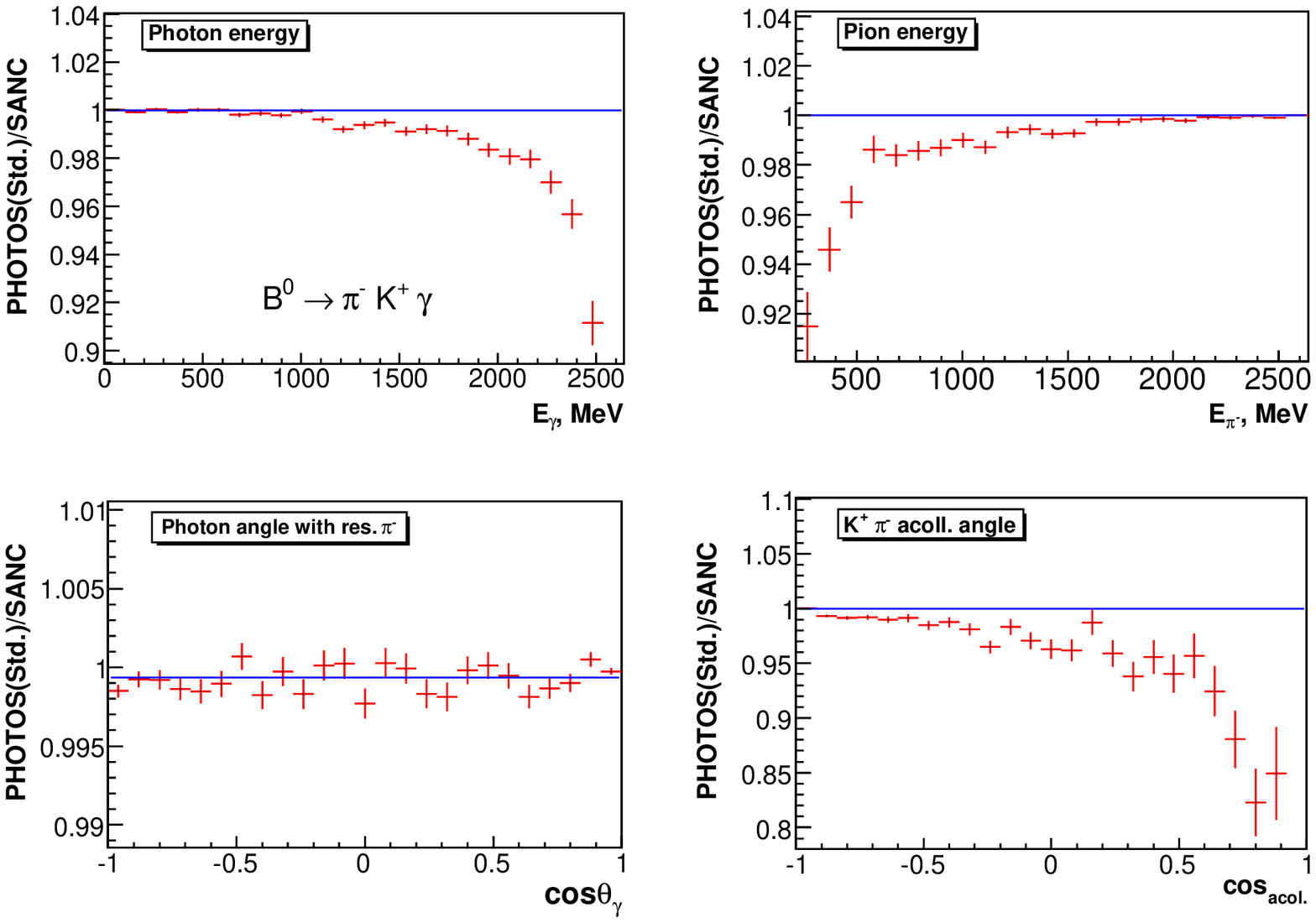}
 \end{center}
 \caption{\label{PmKp_ratio_NotCorrected_p} Results from PHOTOS,
standard version, and SANC for ratios of the $B^0 \to \pi^- K^+(\gamma)$  distributions in 
fig.\ref{PmKp_distr_NotCorrected_p} 
are presented. 
Differences between PHOTOS and SANC are small, but are clearly visible now.
}
\end{figure}

\clearpage

\begin{figure}[b]
\begin{center}
   \includegraphics[ width=160mm,height=270mm, keepaspectratio]{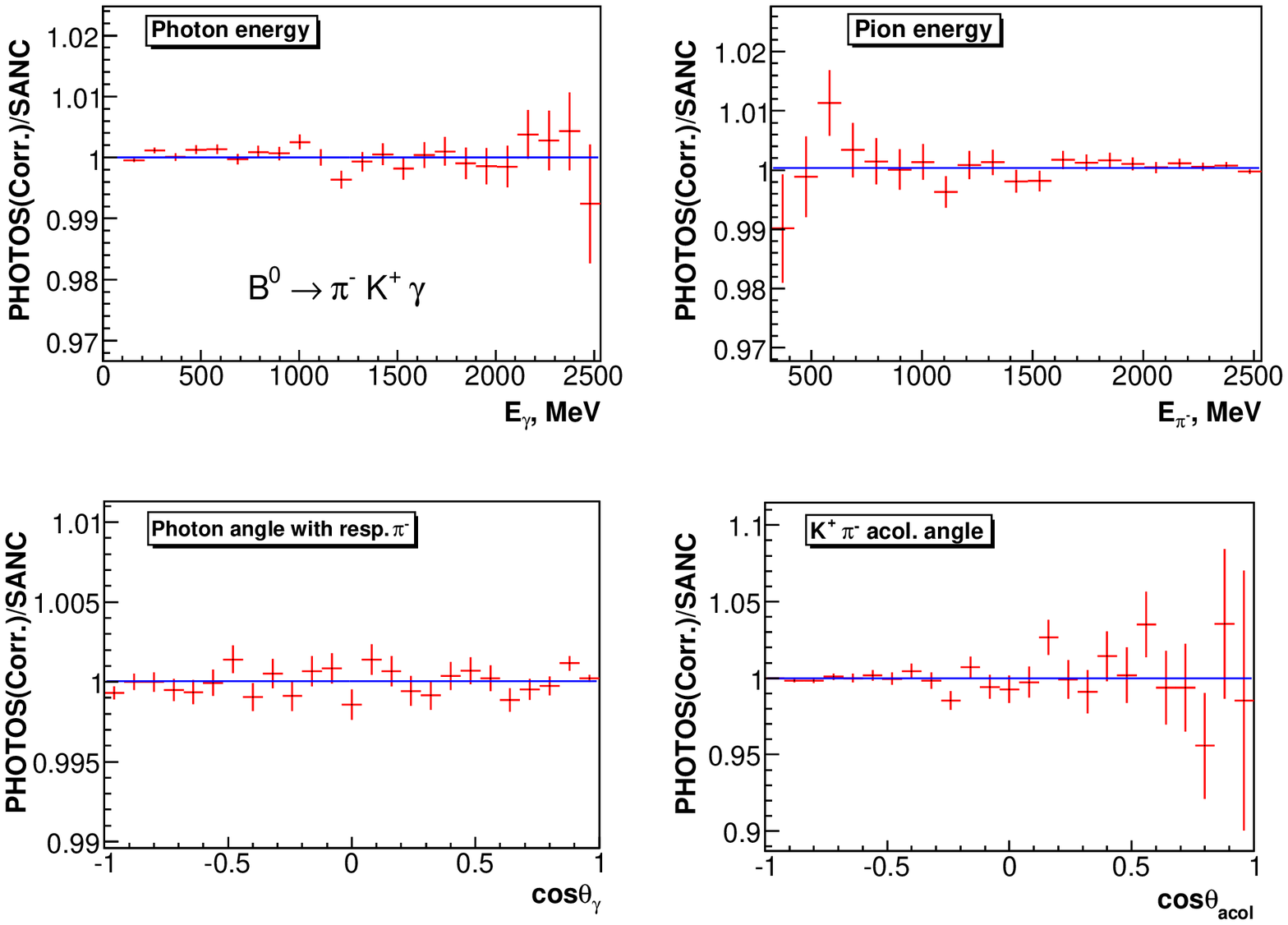}
\end{center}
\caption{\label{PmKp_ratio_corrected_p} Results from PHOTOS with the exact matrix element, and SANC
 for ratios of the  $B^0 \to \pi^- K^+(\gamma)$ distributions. 
Differences between PHOTOS and SANC are below statistical error for samples of $10^9$ events.
}
\end{figure}

\clearpage

\begin{table}[h]
\begin{center}
\begin{tabular}{|r|r|r|r|r|r|r|}
\hline
Channel   & $ \mu_{_{UV}}$  & $E_{test}$  & SANC
                                   & PHOTOS:
  &    &  \\
 & [MeV] & [MeV]  & 
                                   & ${\cal O}(\alpha)$
  &${\cal O}(\alpha^2)$   & ${\cal O}(exp)$\\
\hline
$B^- \to \pi^-\pi^0$ &2500 & 2.6  &0.9291 &  0.9289 & 0.9314 &0.9311 \\  
$B^- \to \pi^-\pi^0$ &2500 & 26   &0.9571 &  0.9569 & 0.9578 &0.9577 \\  
$B^- \to \pi^-K^0$   &2500 & 2.6  &0.9294 &  0.9292 & 0.9318 &0.9314 \\  
$B^- \to \pi^-K^0$   &2500 & 26   &0.9574 &  0.9572 & 0.9580 &0.9580 \\  
\hline
$B^- \to K^-\pi^0$   &2500 & 2.6  &0.9627 &  0.9628 & 0.9636 &0.9634 \\  
$B^- \to K^-\pi^0$   &2500 & 26   &0.9777 &  0.9777 & 0.9779 &0.9779 \\  
$B^- \to K^- K^0$    &2500 & 2.6  &0.9629 &  0.9631 & 0.9639 &0.9638 \\  
$B^- \to K^- K^0$    &2500 & 26   &0.9779 &  0.9779 & 0.9782 &0.9781 \\  
\hline
\hline
$B^0 \to \pi^-\pi^+$ &900 & 2.6  &0.8311  &  0.8306 & 0.8451 &0.8433 \\  
$B^0 \to \pi^-\pi^+$ &900 & 26   &0.8978  &  0.8972 & 0.9019 &0.9016 \\  
$B^0 \to \pi^-K^+$   &900 & 2.6  &0.8662  &  0.8660 & 0.8754 &0.8741 \\  
$B^0 \to \pi^-K^+$   &900 & 26   &0.9193  &  0.9188 & 0.9219 &0.9219 \\  
\hline
$B^0 \to K^-\pi^+$   &900 & 2.6  &0.8661  &  0.8659 & 0.8753 &0.8743 \\  
$B^0 \to K^-\pi^+$   &900 & 26   &0.9193  &  0.9191 & 0.9220 &0.9219 \\  
$B^0 \to K^- K^+ $   &900 & 2.6  &0.9011  &  0.9014 & 0.9066 &0.9057 \\  
$B^0 \to K^- K^+ $   &900 & 26   &0.9407  &  0.9407 & 0.9424 &0.9422 \\  
\hline
\end{tabular}\vspace{0.3cm}
\end{center}
\caption{\it Benchmark results for B decays into pair of scalars: electromagnetic
cumulative of decay width $\Gamma(E_{test})/\Gamma^{\rm Total}$, where  $E_{test}$ 
denotes the maximal  energy which can be carried out by
photons. 
The following input parameters were used: 
$m_B=5279$ MeV, $m_{\pi^0}=135$ MeV, $m_{\pi^\pm}=139$ MeV,  $m_{K^0}=494$ MeV, $m_{K^\pm}=498$ MeV.  
Our results differ  negligibly between
standard PHOTOS and the one with exact matrix element that is why 
only one set of numerical results is provided. For each decay channel
PHOTOS results of first, second and   multiple photon radiation 
 are to a good precision in the following  proportion
 $1-x : 1 - x + x^2/2 : exp(-x)$, where $x$ for each line of the Table
is different; it depends on the decay channel
 and $E_{test}$.  To produce results for our table
samples of $10^7$ events were used. Statistical errors are thus at the level of the 
last significant digit for all the table entries.}
\end{table}


\vspace{.5cm}
\begin{center}
{\bf Appendix: Details and properties  of the explicit phase space parametrization}
\end{center}
\vspace{.3cm}

The formule  (\ref{Jacobians},\ref{barred}) are  central in the definition  
of the PHOTOS algorithm and its  phase-space parametrization. 
For the general 
description details were not important. In practice, they are nonetheless
essential, and all angles masses (or energies) used in formulas (\ref{G-1},\ref{G-2})
must be specified. In particular all reference frames used in the definition of
angles must be defined.

We will start with the detailed description of parametrizations for two 
body  and three body phase spaces, the latter one with an additional 
single photon which accompanies the final state of two massive 
(not necessarily of 
equal masses) objects. In both cases, the decay of an object of the mass $M$ and four 
momentum $P$ is taken into account. The straigtforward extension for the parametrization
of the multibody decay will be introduced with the help of the 
footnote,  properties will be discussed later in the text.

Our particular choice of the phase-space parametrization is of course motivated
by the necessity to regularize the infrared and collinear singularities. 
On the other hand, the definition itself does not need singularities to be 
exposed,
or even to be present at all. To ease reading, let us point out  
 that (at first) we will expect the collinear
singularity  to be present only when the photon becomes parallel to
the direction of  $k_1$. Later, we will discuss  the case when both  final states 
(of momentum $k_1$ and $k_2$) are charged and thus the singularity may appear along the two 
directions.
We will continue with  the case when the photon accompanies a
multi-particle/multi-charge final state
and on the necessity to introduce several simultaneous parametrizations,
to be used  in Monte Carlo parallel generation {\it channels}, used
at each step of iteration as defined in formula (\ref{barred}).

\noindent
In the following eight points we define the angles used for the two-body phase-space parametrization, 
and continue with the definition of phase-space variables of the two-body  plus photon case.
\begin{enumerate}
\item
For the definition of coordinate system in the $P$-rest frame the  $\hat x$ and $\hat y$ axes 
of  the laboratory frame boosted to the rest frame of $P$ can be used. The orthogonal
right-handed system  can be constructed with their help in a standard way.
\item
We choose polar angles $\theta_1$ and $\phi_1$ defining the orientation 
of the four momentum  $\bar k_2$ in the rest frame of $P$. In that frame
$\bar k_1$ and $\bar k_2$ are back to back\footnote{In the 
case of phase space construction
for multi-body decays $\bar k_2$ should read as a state representing the
sum of all decay products of $P$ but $\bar k_1$.}, see fig.~(\ref{pierwsza}).
\item The previous two points would complete the definition of the two-body phase space, 
if  both $\bar k_1$ and $\bar k_2$  
had no measurable  spin  degrees of freedom visualizing themselves 
e.g. through correlations of the secondary decay products' momenta.
Otherwise we  need to know an additional angle $\phi_X$ to complete the set of  Euler
angles defining the relative orientation of the axes of the $P$ rest-frame 
system with the coordinate system
used in the rest-frame of $\bar k_2$ (and possibly also of $\bar k_1$), see fig.~(\ref{druga}).
\item
 If both rest-frames of $\bar k_1$ and $\bar k_2$ are of interest,   
their coordinate systems  are  oriented with respect to $P$
with the  help of   $\theta_1$, $\phi_1$,  $\phi_X$. We  assume 
that the coordinate systems of $\bar k_1$ and $\bar k_2$ are connected by a boost along 
the $\bar k_2$ direction, and in fact share axes: $  z'  \uparrow \downarrow  z''$,
 $ x' \uparrow \uparrow  x''$,  $y' \uparrow \downarrow y''$.
\item 
Let us turn now to the three-body phase space parametrization. 
We take the photon energy $k_{\gamma}$  in the rest frame 
of $P$, with its help we calculate: photon, $k_1$ and $k_2$  energies,
all  in $k_1+k_2$ frame.

\item We use the angles $\theta$, $\phi$,  in the 
rest-frame of the $k_1+k_2$ pair: angle $\theta$ is  an angle between the photon and 
$k_1$ direction (i.e. $-  z''$ ).  
Angle $\phi$ defines the photon azimuthal angle around $ z''$,
 with respect to $ x''$ axis (of the  $k_2$ rest-frame), 
see fig.~(\ref{trzecia}).

\item If all $k_1$,  $k_2$ and $k_1+k_2$ rest-frames
 exist, then the  $x$-axes for the three frames  are chosen to  coincide. 
It is possible, because 
they are all connected by the boosts along the common $ z''$ direction,
see fig.~(\ref{trzecia}). The axes of $k_1+k_2$ rest-fame are not drawn 
explicitly.

\item To define orientation of $k_2$ in P rest-frame
coordinate system, and to complete construction
of the whole event,
we will re-use Euler angles of $\bar k_2$:  $\phi_X$, $\theta_1$ and $\phi_1$
(see figs. \ref{czwarta} and  \ref{piata}),  defined again of course 
in the rest frame of $P$.

\end{enumerate}

 That completes our definitions of parametrizations for the two-body and three-body phase spaces
necessary to  define transformation $G$ 
of  formulas (\ref{G-1},\ref{G-2}).
Before commenting on the properties of our parametrizations, let us note that 
these parametrizations were already used  and defined 
in ref.~\cite{Barberio:1994qi},
in all details, except for  the function  of 
(only implicitly introduced there) angle $\phi_X$. 
For some readers,  definition from that paper may be also easier to follow.
  
Let us  comment, now, on these 
 properties of our parametrizations which are important
for construction of the PHOTOS algorithm.
\begin{enumerate}
\item [a)]
The parametrizations of the two-body and  three-body  phase-space
(photon included) are  used for the explicit
kinematical construction denoted by formula (\ref{Jacobians}). 
We can replace the roles played by $k_1$ and $k_2$. This simple operation
 leads to 
a new phase-space parametrization, which can be used in a
second branch of the Monte Carlo generation.
\item[b)]
The  phase-space Jacobians  (factor $W^{n+1}_n$  
of  (\ref{Jacobians}))
are identical for the 
two branches; this factor is also never larger than  $k_\gamma {1 \over 2(2\pi)^3}$.
\item[c)] Angle $\theta$ of the first branch  coincides with $\pi-\theta$ of the second one.
\item[d)]  In the soft ($k_\gamma \to 0$) and collinear 
($\theta \to 0$ or $\pi$) limits, angles $\theta_1$, $\phi_1$, $\phi_X$ of the 
two branches converge to each other 
(in these limits they may differ by $\pi$ or $2\pi$). 
\item[e)]
Properties (c) and (d) are convenient for  our construction of the weights
given by formula (\ref{wgt1}), because they coincide with the 
similar properties of the exact matrix element. 
\item[f)] Thanks to property (b), also the first version of (\ref{wgt1}) 
is exact. In fact, this first version is more suitable for multi-photon 
radiation, if first order matrix element is used only. This observation 
required comparisons with second order matrix elements \cite{Golonka:2006tw}.
The choice of $\bar k_2$  (or $k_2$) direction  to define 
 $\theta_1$, $\phi_1$, rather than  $\bar k_1$, 
was also motivated by the properties  
of the decay matrix elements.
\end{enumerate}

Let us present now some further observations\footnote{  Note, that the approximations to be discussed in the following points,
result from  matching kinematical branches and  
affect the way how phase-space Jacobians are used in (\ref{wgt1}). The full phase 
space remain  covered, as is the  case of formule (\ref{Jacobians},\ref{barred}) denoting exact 
phase-space parametrization of single and multi-photon final state.
}, which go beyond
NLO corrections 
 for the processes 
discussed in ref.\cite{Golonka:2006tw} and  in the present paper,
and point 
to applications for a multi-body/multi-charge final states.

\begin{itemize}
\item Property (d) extends to more than  two-body decays, and also to 
 cases when there are  more than two charged
particles present in the final state. The relation between angles
  $\theta_1$, $\phi_1$, $\phi_X$ of the distinct branches is more
complex, but in the discussed limits still independent from  $\theta$ and
 $\phi$.
\item
Extended property (d) and property (e) enable the use of   (\ref{wgt1}) for  multi-photon 
radiation; this also holds in the case when more than two charged particles 
are present in the final state.
\item That is why, in the case of two-body decays (plus bremsstrahlung photons),  
such type of phase-space treatment is sufficient for the  NLO
 precision. 
\item
For the  NNLO precision, in 
matching of the two mappings for the collinear singularities\footnote{
Such matching is necessary
for the two branches of the generation, used  to presample 
collinear singularities along the directions  
of $k_1$ and $k_2$, to be used simultaneously in construction of each
event.}
 another factor of the type 
$\lambda^{1/2}(...) / \lambda^{1/2}(...)$ would have to be  included
  in  $W^{n+1}_n$ of  formulas (\ref{Jacobians},\ref{wgt1}). In fact in such a case 
the exact 
 multi-photon phase space parametrization 
would be preserved.
\item For each additional charged decay product present in the final state,
still another factor of the type 
$\lambda^{1/2}(...) / \lambda^{1/2}(...)$ is needed in  $W^{n+1}_n$ to assure multichannel generation 
with the exact treatment of the phase space. 
\item Even without future refinements (as explained in the previous two 
points)  our phase space parametrization  is sufficient for  NLO and NLL 
precision  for the two-body (two-charges)  decays, accompanied
with arbitrary number of photons. In a general case, when more 
than two charged particles are present in final state,  
such  phase space  parametrization remains sufficient
for  LL only, even though also in this case, the  full multi-photon phase 
space is covered.
At present, the resulting precision is sufficient and does not justify those
easy changes.
\item In our choice of phase space parametrization (point 1), we have dropped 
some  details, the choice 
of $\hat x, \; \hat y, \; \hat z$ axes of the $P$ rest-frame were not specified.
Indeed, for the decay of a scalar object, such 
as that discussed in the present paper, every choice is equivalent. 
In general, it is not the case. Already in case of the $Z$ boson decay, the 
choice of the $\hat z$ axis parallel to the direction of the incoming beam 
of the same charge as $k_2$ is advantageous,  see ref. \cite{Golonka:2006tw}
where the process  $e^+e^- \to Z \to l^+ l^- n(\gamma)$ was studied. 
In this case the direction of the incoming beam coincides with  
the spin state of  $Z$, and the choice  simplify expression for matrix element.

\item Finally, we may consider extending our method beyond the decays.
In the case of the $t$-channel processes, or 
initial state radiation, particular choices for the frames and 
angles would be also essential for the
construction to match the structure of the matrix element singularities. 
Some hints into that direction can be seen already now, refs. 
\cite{Berends:1982ie,Was:2004ig}.
 \end{itemize}

\begin{figure}[!h]
  \begin{center}
    \includegraphics[ width=100mm,height=130mm, keepaspectratio]{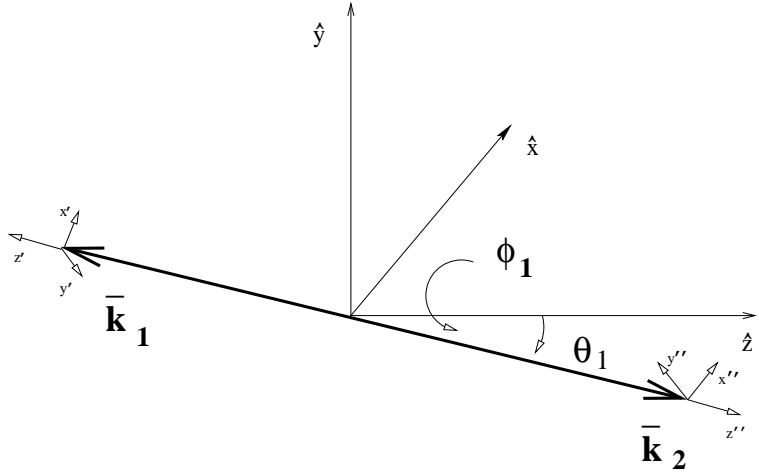}
  \end{center}
  \caption{\label{pierwsza} The angles $\theta_1$, $\phi_1$ 
defined in the rest-frame of $P$ 
and used in parametrization of two-body phase-space.  
  }
\end{figure}

\begin{figure}[!h]
  \begin{center}
    \includegraphics[ width=100mm,height=130mm, keepaspectratio]{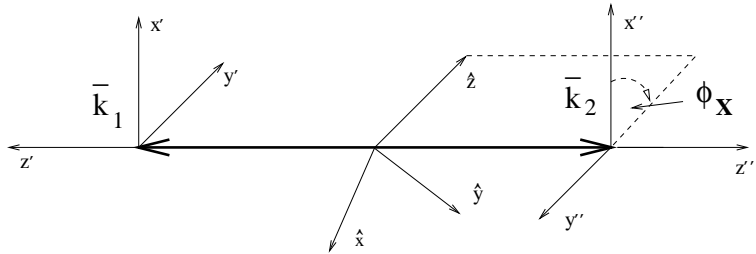}
  \end{center}
  \caption{\label{druga} Angle $\phi_X$ is also defined in the rest-frame
of $P$ as an angle between (oriented) planes spanned on: (i) $\bar k_1$ and
$\hat z$-axis of the $P$ rest-frame system, and (ii)   $\bar k_1$ and
$x''$-axis of the  $\bar k_2$ rest frame. It completes definition of the phase-space 
variables if internal orientation of $\bar k_1$ system is of interest. In fact,
Euler angle $\phi_X$ is inherited from unspecified in details, parametrization
of  phase space used to describe possible
future decay of $\bar k_2$ (or $\bar k_1$).
  }
\end{figure}

\begin{figure}[!h]
  \begin{center}
    \includegraphics[ width=100mm,height=130mm, keepaspectratio]{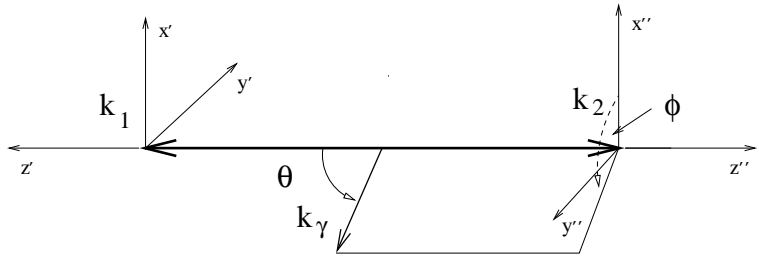}
  \end{center}
  \caption{\label{trzecia} The angles $\theta$, $\phi$ are used to construct
the four-momentum of $k_\gamma$ in the rest-frame of $k_1+k_2$ pair (itself 
not yet oriented with respect to $P$ rest-frame). To calculate energies of 
$k_1$, $k_2$ and  photon, it is enough to know $m_1$,  $m_2$,  $M$ and
photon energy  $k_\gamma$ of the $P$ rest-frame.
  }
\end{figure}

\begin{figure}[!h]
  \begin{center}
    \includegraphics[ width=100mm,height=130mm, keepaspectratio]{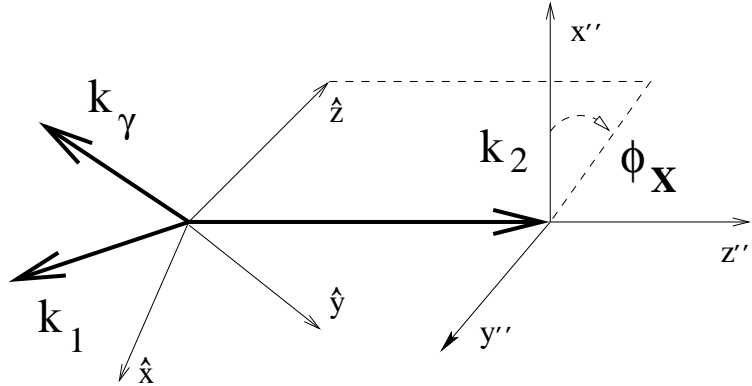}
  \end{center}
  \caption{\label{czwarta} Use of angle $\phi_x$ in defining orientation of 
$k_1$, $k_2$ and photon in the rest-frame of $P$. At this step only the 
plane spanned on $P$ frame axis $\hat z$ and $k_2$ is oriented with respect to $k_2\times  x''$ plane.
  }
\end{figure}

\begin{figure}[!h]
  \begin{center}
    \includegraphics[ width=100mm,height=130mm, keepaspectratio]{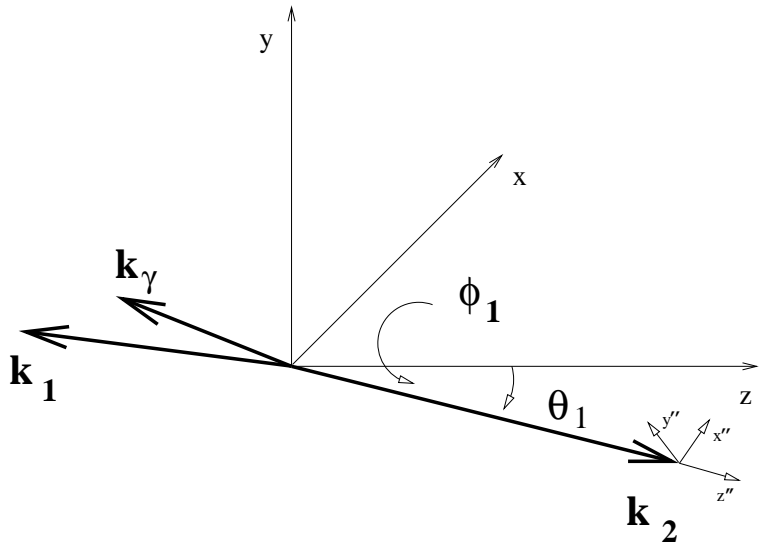}
  \end{center}
  \caption{\label{piata} Final step in event construction. Angles $\theta_1$,
$\phi_1$ are used. The final orientation of $k_2$ coincide with this of
$\bar k_2$.
  }
\end{figure}

\clearpage

\providecommand{\href}[2]{#2}

\end{document}